\documentclass{aa}
\usepackage{graphicx}
\usepackage{tabularx}
\usepackage{subcaption}
\usepackage{tablefootnote} 
\usepackage{enumitem} 
\usepackage{amsmath}
\usepackage{multirow} 
\usepackage{mathrsfs}
\usepackage{txfonts}
\usepackage{hyperref}
\usepackage{cleveref}
\usepackage{footnote}
\usepackage{fancyhdr}
\makesavenoteenv{table}

\newcommand{\RM}{\text{RM}}
\newcommand{\GRM}{\text{GRM}}
\newcommand{\RRM}{\text{RRM}}
\providecommand{\abs}[1]{\lvert#1\rvert} 
\usepackage[dvipsnames]{xcolor}
\usepackage{placeins}
\usepackage{orcidlink}

\usepackage{natbib}
\bibpunct{(}{)}{;}{a}{}{,}
\hypersetup{colorlinks=true,
            linkcolor=blue,
            citecolor=blue,
            filecolor=magenta,
            urlcolor=blue
            }
\setcitestyle{authoryear}

\begin{document}

   \title{Magnetic Fields in the Shapley Supercluster Core with POSSUM: Challenging Model Predictions}

   \author{D. Alonso-L\'opez \orcidlink{0000-0001-9006-0725}
          \inst{1}
          \and
          S. P. O'Sullivan \orcidlink{0000-0002-3968-3051} \inst{1}
          \and 
          A. Bonafede \orcidlink{0000-0002-5068-4581}  \inst{2,3}
          \and
          L. M. Böss \orcidlink{0000-0003-4690-2774} \inst{4}
          \and
          C. Stuardi \orcidlink{0000-0003-1619-3479} \inst{3}
          \and
          E. Osinga \orcidlink{0000-0002-5815-8965} \inst{5}
          \and
          C. S. Anderson \orcidlink{0000-0002-6243-7879} \inst{6}
          \and
          C. L. Van Eck \orcidlink{0000-0002-7641-9946} \inst{6}
          \and
          E. Carretti \orcidlink{0000-0002-3973-8403} \inst{3}
          \and 
          J. L. West \orcidlink{0000-0001-7722-8458} \inst{7}
          \and
          T. Akahori\inst{8}
          \and
          K. Dolag\inst{9, 10}
          \and
          S. Giacintucci\inst{11}
          \and
          A. Khadir\inst{12, 5}
          \and
          Y. K. Ma \orcidlink{0000-0003-0742-2006} \inst{13}       
          \and 
          S. Malik \orcidlink{0000-0003-4147-626X} \inst{1}
          \and
          N. McClure-Griffiths \orcidlink{0000-0003-2730-957X} \inst{6}
          \and
          L. Rudnick\inst{14}
          \and
          B. A. Seidel \orcidlink{0009-0000-3688-4379} \inst{9}
          \and
          S. Tiwari\inst{15}
          \and
          T. Venturi\inst{3} }

   \institute{
   Departamento de F\'isica de la Tierra y Astrof\'isica $\&$ IPARCOS-UCM, Universidad Complutense de Madrid, 28040 Madrid, Spain \email{dalons07@ucm.es}
   \and Dipartimento di Fisica e Astronomia, Università degli Studi di Bologna, via P. Gobetti 93/2, 40129 Bologna, Italy 
   \and INAF – Istituto di Radioastronomia, via P. Gobetti 101, 40129 Bologna, Italy
   \and Department of Astronomy and Astrophysics, The University of Chicago, IL 60637, USA
   \and Dunlap Institute for Astronomy and Astrophysics, University of Toronto, 50 St. George Street, Toronto, M5S 3H4, ON, Canada
   \and Research School of Astronomy $\&$ Astrophysics, The Australian National University, Canberra ACT 2611, Australia
   \and National Research Council Canada, Herzberg Research Centre for Astronomy and Astrophysics, Dominion Radio Astrophysical Observatory, PO Box 248, Penticton, BC V2A 6J9, Canada
   \and Mizusawa VLBI Observatory, National Astronomical Observatory of Japan, 2-21-1, Osawa, Mitaka, Tokyo 181-8588, Japan
   \and Universitäts-Sternwarte, Fakultät für Physik, Ludwig-Maximilians-Universität München, Scheinerstr.1, 81679 München, Germany
   \and Max-Planck-Institut für Astrophysik, Karl-Schwarzschild-Str. 1, 85748 Garching, Germany
   \and U.S. Naval Research Laboratory, 4555 Overlook Ave. SW, Washington, DC 20375, USA
   \and David A. Dunlap Department of Astronomy and Astrophysics, University of Toronto, 50 St. George Street, Toronto, M5S 3H4, ON, Canada
   \and Max-Planck-Institut f\"ur Radioastronomie, Auf dem H\"ugel 69, 53121 Bonn, Germany
   \and Minnesota Institute for Astrophysics, University of Minnesota, 116 Church Street SE, Minneapolis, MN 55455, USA
   \and Trottier Space Institute, McGill University, 3550 University Street, Montreal, QC Canada, H3A 2A7 
   }

   \date{Received 7 July 2025; Accepted 17 November 2025}

  \abstract 
   {Faraday Rotation Measure (RM) Grids provide a sensitive means to trace magnetized plasma across a wide range of cosmic environments.}
   {We study the RM signal from the Shapley Supercluster Core (SSC), in order to constrain the magnetic field properties of the gas. The SSC region consists of two galaxy clusters A3558 and A3562, and two galaxy groups between them, at $z\simeq  0.048$. }
   {We combine RM Grid data with thermal Sunyaev-Zeldovich effect data, obtained from the POlarisation Sky Survey of the Universe’s Magnetism (POSSUM) pilot survey, and Planck, respectively. 
   To robustly determine the gas density, its magnetic field properties, and their correlation $\abs{\Vec{B}}\propto n_e^{\eta}$, we study the $\RM$ scatter in the SSC region ($\mathfrak{S}_{\RM}$) and its behavior as a function of distance to the nearest cluster/group $(d_{\text{nrst}})$. We compare observational results with semi-analytic Gaussian random field models and more realistic cosmological MHD simulations.}
   {With a sky-density of 36 RMs/deg$^{2}$, we detect an excess RM scatter of $30.5\pm 4.6 \,\, \mathrm{rad/m^2}$ in the SSC region. Comparing with models, we find an average magnetic field strength of $\sim$1-3 $\mu$G (in the groups and clusters). 
   The $\mathfrak{S}_{\RM}(d_{\text{nrst}})$ profile, derived from data ranging from $\sim$0.3-$1.8\,r_{500}$ for all objects, is systematically flatter than expected compared to models, with $\eta<0.5$ being favored. Despite this discrepancy, we find that cosmological MHD simulations matched to the SSC structure most closely align with scenarios where the magnetic field is amplified by the turbulent velocity ($v_{\text{turb}}$) in the intercluster regions $B_{\mathcal{F}}\propto n_e^{1/2} v_{\text{turb}}$ on scales $d_{\text{nrst}}\lesssim 0.8$. }
   {The dense RM grid and precision provided by POSSUM allows us to probe magnetized gas in the SSC clusters and groups on scales within and beyond their $r_{500}$.  
   Flatter-than-expected $\RM$ scatter profiles reveal a significant challenge in reconciling observations with even the most realistic predictions from cosmological MHD simulations in the outskirts of interacting clusters. }


   \maketitle
   
    \begingroup
    \renewcommand{\thefootnote}{}
    \footnotetext{IPARCOS-UCM-25-038}
    \endgroup
\section{Introduction} \label{sect:intro}

Magnetic fields are known to be ubiquitous in the Universe. They span a wide range of scales and strengths: from stars and magnetars (up to $\sim10^{15}\,\mathrm{G}$) \citep{magnetars2025} all the way to $\sim\mathrm{nG}$ in cosmic web filaments \citep{carrettiNatureLOFARRotation2025}. In the overdensities of the cosmic web, i.e., superclusters, galaxy clusters, large-scale magnetic fields have also been found. The turbulent intracluster medium (ICM) is permeated by $\sim \mu\mathrm{G}$ magnetic fields at $\sim 1$-$100\,\mathrm{kpc}$ scales, see \citet{govoniMagneticFieldClusters2004, Donnert2018}, for reviews of cluster magnetic fields. These magnetic fields play important roles on the dynamics and understanding of the baryonic content of the structures they permeate.

The baryonic content of the local Universe is well constrained both by the predictions of Big Bang Nucleosynthesis through the analysis of the abundance of primordial light elements and observations of the Lyman-$\alpha$ forest at high redshift \citep{nicastroMissingBaryonsWarmHot2008}, and Cosmic Microwave Background \citep{aghanimPlanck2018Results2020}. However, only $\sim 70\%$ of these baryons have been found in observations of the local Universe \citep{shullBaryonCensusMultiphase2012}. Nonetheless, the remaining $\sim 30\%$, the so-called, missing baryons, are now claimed to be accounted for, although the uncertainties are yet to be lowered for a more definitive conclusion \citep{driverChallengeMeasuringMapping2021}. They are expected to lie in the Warm Hot Intergalactic Medium (WHIM) \citep{nicastroObservationsMissingBaryons2018a, macquartCensusBaryonsUniverse2020}. The WHIM is made up of diffuse ionised gas filling the filaments of the cosmic web, thus, these missing baryons lie in low density regions, making them hard to detect. 
Cosmic web filaments and intercluster bridges are such low density structures in the Universe where these baryons are expected be. Recently, evidence has been found for magnetic fields in these filaments \citep{carrettiNatureLOFARRotation2025} and it is therefore natural to think that intercluster bridges, where diffuse radio emission, and thermal Sunyaev-Zeldovich (tSZ) effect signals have been detected \citep{pignataroMindGapA20612024, pignataroProbingDiffuseRadio2024, tanimuraDetectionInterclusterGas2019, graaffProbingMissingBaryons2019, radiconiThermalNonthermalComponents2022}, are magnetized as well. It is then interesting to determine their level of magnetization and their magnetic field strengths and properties.

The Shapley Supercluster Core (SSC) ($z\simeq0.048$) contains one such intercluster bridge between the massive galaxy clusters Abell 3558 and Abell 3562 (hereafter A3558, A3562), detected by Planck through the tSZ effect \citep{aghanimPlanck2015Results2016}. The tSZ emission of two massive groups of galaxies was also detected in this bridge (SC 1327 and SC 1329). The masses $M_{500}$ and radii $r_{500}$\footnote{Where $r_{500}$ is the radius of the sphere that contains a density equal to $500\rho_c(z)$ where $\rho_c(z)$ is the critical density of the Universe at the redshift of the object in consideration, and $M_{500}$ the mass associated with this overdensity.} of these four objects can be found in Table \ref{tab:m500_r500_obs}. The projected distance between the centers of the Abell clusters is $\simeq 4.2\,\mathrm{Mpc}$. Such a system provides a unique laboratory for the study of the diffuse magnetized gas in intercluster regions to derive the properties of their magnetic field through Faraday Rotation. 

\begin{table}[h]
\caption{$M_{500}$ and $r_{500}$ of the four objects comprising the Shapley Supercluster Core (SSC).  }
\centering
\renewcommand{\arraystretch}{1.2}
\begin{tabular}{c|cccc}
Mass/Radius & A3558 & A3562 & SC 1327 & SC 1329 \\ \hline \hline
$M_{500}$ [$10^{14}M_\odot$] & 9.8 & 4.4 & 2 & 0.5 \\
$r_{500}$ [Mpc] & 1.5 & 1.2 & 0.9 & 0.6 
\end{tabular}
\tablefoot{The masses were obtained from \citet{venturiRadioFootprintsMinor2022}, while we scaled the $r_{200}$ in \citet{higuchiShapleySuperclusterSurvey2020} using $r_{500}\approx 0.65\,r_{200}$ \citep{reiprich2013}.}
\label{tab:m500_r500_obs}
\end{table}

Linearly polarized light that travels through an ionised and magnetized medium experiences Faraday Rotation, i.e., a rotation of the linear polarization vector $\chi$, which is proportional to the square of its wavelength: $\Delta\chi = \RM \cdot\lambda^2$, where the constant of proportionality is defined as the rotation measure ($\RM$), e.g., \citet{ferriereCorrectSenseFaraday2021}. Since the $\RM$ is an integral along the line of sight, the quantity we measure is the sum of all the contributions from every Faraday screen $\mathcal{S}_i$
\begin{equation}
    \RM_{\text{obs}} = \sum_i\RM_i = \frac{e^3}{2\pi m_e^2c^4}\sum_i\int_{\mathcal{S}_i} n_e(l)\,B_{\parallel}(l)\, dl.
\label{eq:RMobs}
\end{equation}
The integral goes from the source to the observing point. The constants are: $e$ the absolute value of the charge of the electron, $m_e$ its mass and $c$ the speed of light. The $\RM$ depends on the product of the number density of thermal free electrons ($n_e$) and the magnetic field component along the line of sight ($B_{\parallel}$)\footnote{Note that we neglect the effect of redshift dilution \citep{Akahori11}. The on-off methodology in Section \ref{sect:sigma_RM_on_target}, Eqn. \eqref{eq:signal_obs} ensures that this assumption is valid, given that we don't expect major differences between the on and off-target redshift distributions. Furthermore, at the redshift of the SSC ($z\approx 0.048$), the factor $(1+z)^{-2}$ is of order 1. }. 
Therefore, to properly quantify how much of the observed rotation is actually due to the Shapley Supercluster, we need to properly account for all other possible contributions, typically local contributions to the source itself, intervening extragalactic structures along the way, and most importantly the Galactic contribution, denoted by $\GRM$. When analysing $\RM$ data we will consider differences in the $\RM$ dispersion of different sub-samples. Thus, any contribution to the dispersion that can be set as common to both sub-samples, such as the intrinsic contribution from the sources, will cancel out. Given that the $\GRM$ contribution can be different at different scales throughout the Shapley region we aim to study, it is necessary to model it in detail. \\

Radio polarization data has been widely used to study magnetic fields in clusters through Faraday Rotation \citep{murgiaMagneticFieldsFaraday2004, govoniIntraclusterMagneticField2006, guidettiIntraclusterMagneticField2008, bonafedeComaClusterMagnetic2010, govoniRotationMeasuresRadio2010, vaccaIntraclusterMagneticField2012, govoniSardiniaRadioTelescope2017, stuardiIntraclusterMagneticField2021}. However, one of the main sources of uncertainty in these $\RM$ grid studies, is the limited number of polarized sources whose line-of-sight (l-o-s) go through the given object of study \citep{johnsonCharacterizingUncertaintyCluster2020}. This limited sampling of the density and magnetic field information is one of the main drivers for the Australian Square Kilometer Array Pathfinder-Polarisation Sky Survey of the Universe's Magnetism (ASKAP-POSSUM) collaboration to build the densest RM grid of the southern sky to date, with about 1 million expected extragalactic polarized sources, aiming for $\sim30$-$50\,\RM/\mathrm{deg^2}$ \citep{gaenslerPolarisationSkySurvey2025}. The previous densest wide-area catalog has $\sim1\,\RM/\mathrm{deg^2}$ \citep{taylorROTATIONMEASUREIMAGE2009}. Recent studies with Square Kilometer Array (SKA) pathfinder telescopes data have shown its capabilities to map and detect ionised and magnetized gas in the outskirts of galaxy clusters \citep{andersonEarlySciencePOSSUM2021, loiMeerKATFornaxSurvey2025} as well as the $\RM$ signature of galaxy groups \citep{andersonProbingMagnetisedGas2024}. Other recent studies of magnetic fields in galaxy clusters have shown the potential of statistical stacking as an alternative to wide area dense $\RM$ grids \citep{osingaDetectionClusterMagnetic2022, osingaProbingClusterMagnetism2025}. 

 In this work, we study the magnetic field properties of the SSC with ASKAP-POSSUM\footnote{\href{https://possum-survey.org/}{https://possum-survey.org/}} data from the POSSUM Pilot II survey. We will focus on the potential of these dense $\RM$ grids to estimate the magnetic properties of the gas in the intercluster bridge region between the clusters A3558 and A3562. The underlying assumed cosmology throughout this paper is a $\Lambda$CDM with $\Omega_m=0.3$, $\Omega_{\Lambda}=0.7$ and $H_0=70\,\mathrm{km/s/Mpc}$. The linear scale at the redshift of Shapley is thus $0.941\,\mathrm{kpc/}$''. Section \ref{sect:data} is dedicated to the data used in this work. Sections \ref{sect:obs_results} and \ref{sect:model_results} show the results derived from the data and the different modeling approaches we followed, respectively. Section \ref{sect:discussion} contains the discussion on our results. In Sect. \ref{sect:conclusions} we conclude our findings. Appendix \ref{sect:obs_methods} and \ref{sect:model_methods} provide ancillary results and outline the methods used to extract information from the observations and the modeling, respectively.

 Throughout this paper, the symbol $\sigma$ will be used as a measure of the scatter in the $\RM$ data, and it will always refer to the Median Absolute Deviation (MAD) standard deviation defined as $\sigma=1.4826\, \text{MAD}$, which is less sensitive to outliers than the traditional standard deviation.

\section{Data}\label{sect:data}

\subsection{Radio polarization and Faraday Rotation data from ASKAP-POSSUM}\label{sect:askap}

\begin{table*}[ht!]
\caption{Summary of observations.}
\centering
\renewcommand{\arraystretch}{1.35}
\begin{tabular}{c|cccccccc}

SBID & Band & Central Freq. [MHz] & Field & RA [hms], Dec [dms] & Obs. Date & Obs. Time & rms [$\mu$Jy/beam] \\ \hline \hline

SB43137 & 1 & 943.5 & Core & 13:29:47, -30:17:24 & 03-Aug-2022 & 10h & 46.98 \\
SB34634 & 1 & 943.5 & South & 13:33:10, -34:55:12 & 22/23-Dec-2021 & 10h & 47.40 \\
SB43206 & 2 & 1151.5 & Core & 13:28:51, -29:52:48 & 06-Aug-2021 & 8h & 57.18 \\
SB43228 & 2 & 1151.5 & Core & 13:30:56, -30:19:48 & 07-Aug-2021 & 8h & 58.83 \\
SB43300 & 2 & 1151.5 & South & 13:33:52, -35:13:48 & 10-Aug-2021 & 8h & 60.82 \\

\end{tabular}
\label{tab:observaciones}
\end{table*}

\begin{figure}[h!]
    \centering
    \includegraphics[width=1.2\linewidth,clip=true,trim=6.5cm 2cm 2cm 4cm]{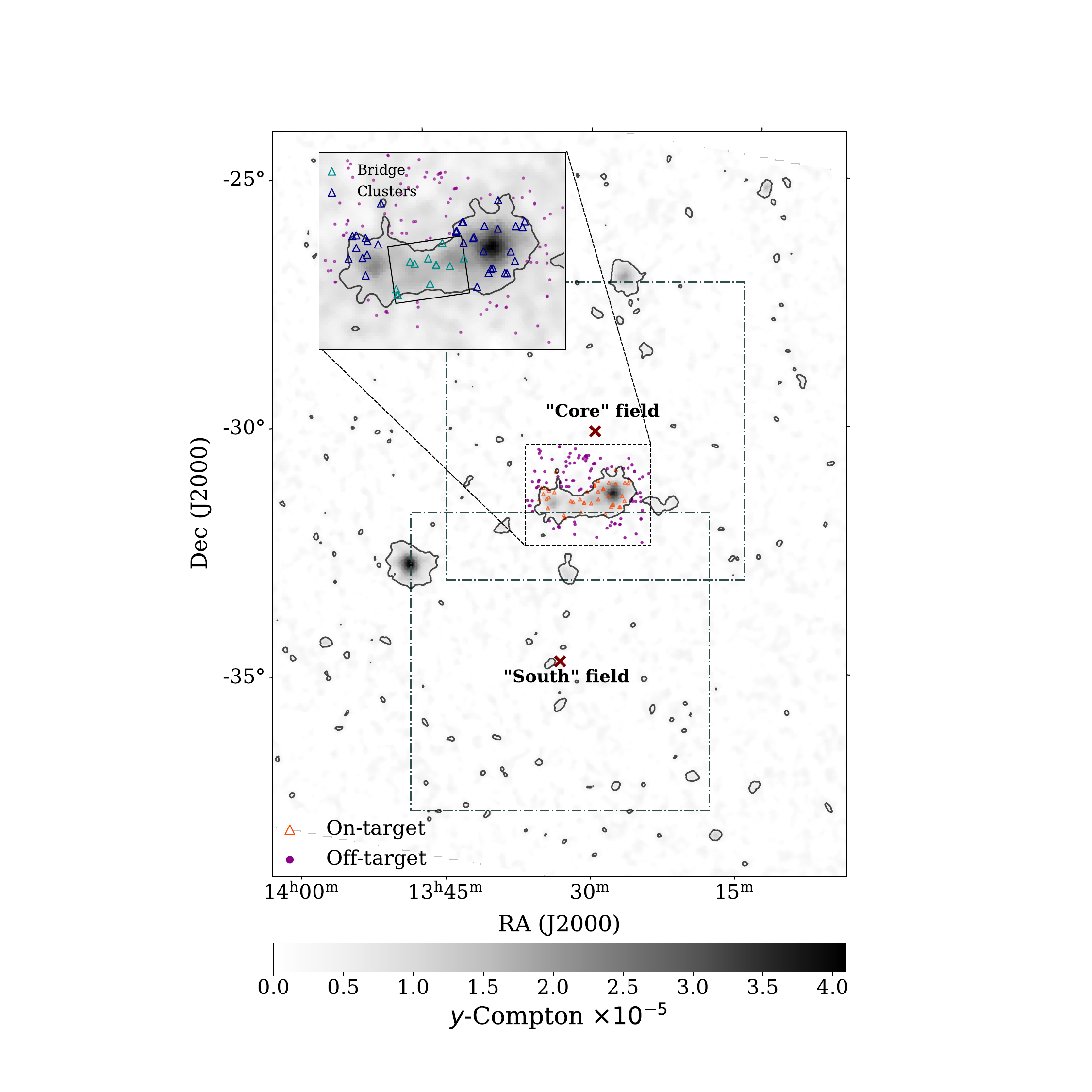}
    \caption{In this figure we aim to show in a clear and visual way the POSSUM $\RM$s used in this work and the definition of the different $\RM$ sub-samples. The background shows the Planck tSZ effect $y$-map. The dash-dotted squares represent the (6 deg)$^2$ Band 1 ASKAP fields, namely, the ``core'' and ``south'' fields, with their centers represented by the red crosses. The black contour represents the $y_{\text{bdry}}=4.24\times 10^{-6}$ value used to define the on-target (orange triangles) and off-target (purple dots) samples. These two are inside the 1.71$\times$2.42 deg$^2$ (5.8$\times$8.2 Mpc$^2$) $y$-map cutout used for the analysis (see Sect. \ref{sect:planck}). The zoomed-in region at the top left corner of the plot shows the bridge (cyan triangles) and clusters (blue triangles) sub-samples, all of which belong in the on-target region. The bridge box used to define the bridge sources is also represented to ease the interpretation of the plot. }
    \label{fig:subsamples}
\end{figure}

In this work we have used observations from the ASKAP-POSSUM pilot survey by combining data from Band 1 ($800$-$1088$ MHz) and Band 2 ($1296$-$1440$ MHz) with a spectral resolution of $1\,\mathrm{MHz}$ \citep{gaenslerPolarisationSkySurvey2025}. The target of these observations was the Shapley Supercluster in two different fields: ``core'' and ``south'' (as shown in Fig. \ref{fig:subsamples}). The tile center coordinates for ``core'' and ``south'' in Band 1 and Band 2 are listed in Table~\ref{tab:observaciones}. 
ASKAP's field of view is $30\,\mathrm{deg^2}$, while for the typical integration time and rms noise consult Table \ref{tab:observaciones}.

\subsubsection{Deriving the RM catalog}

For each field, the Stokes $I$, $Q$ and $U$ image cubes for the fields were obtained from the CASDA repository\footnote{\href{https://data.csiro.au/domain/casdaObservation}{https://data.csiro.au/domain/casdaObservation}}, which were produced by the ASKAP Observatory pipeline using the ASKAPsoft software package \citep{guzmanASKAPScienceData2019}. The on-axis polarization calibration was done for each of the 36 beams (closepack36 footprint with beam pitch of $0.9\,\mathrm{deg}$) using the unpolarized bandpass-corrected primary calibrator (PKS B1934-638).
The off-axis leakage in the Stokes $Q$ and $U$ cubes was corrected using beam models derived from holography observations (which was estimated to result in instrumental polarization levels of order $1\%$ or less of Stokes $I$ for these particular observations by the POSSUM data validation team \citep{vanderwoudePrototypeFaradayRotation2024, gaenslerPolarisationSkySurvey2025}). A Stokes $I$ catalog for each field is also provided by the ASKAP Observatory team using the Selavy software package \citep{whitingSourceFindingAustralianSquare2012, whiting2017}.
In order to quantify the linear polarization and Faraday rotation properties of all Stokes $I$ sources, a development version of the POSSUM Polarimetry Pipeline was used. Initially it was ensured that the image cubes were convolved both spatially and spectrally to a common angular resolution of 20" (using the beamcon\_3D routine from racs\_tools\footnote{\href{https://github.com/AlecThomson/RACS-tools}{https://github.com/AlecThomson/RACS-tools}}).
The $Q$ and $U$ image cubes were then corrected for the average ionospheric Faraday rotation during the observation using the package FRion\footnote{\href{https://frion.readthedocs.io/en/latest/}{https://frion.readthedocs.io/en/latest/}}.
The 1D $I$$Q$$U$ spectra were then extracted at the position of each Stokes $I$ catalog entry. To account for any potential contamination of the Stokes $Q$ and $U$ spectra by diffuse polarized emission from the Milky Way, the median Stokes $Q$ and $U$ emission in a $109''$ annulus surrounding each source was subtracted (see eg. \citet{vanderwoudePrototypeFaradayRotation2024} for details). The optimal annulus size was determined through simulated source tests in an internal POSSUM memo \citep{oberhelman_possum_memo}. The $109''$ outer radius is large enough such that the error on the subtracted median value does not affect the noise in the subtracted spectra, while also being small enough such that the procedure does not subtract diffuse emission that is unrelated to that contaminating the target spectra.. RM synthesis was then applied to the individual spectra using the RM-Tools software package rmsynth\_1d\footnote{\href{https://github.com/CIRADA-Tools/RM-Tools/wiki/RMsynth1D}{https://github.com/CIRADA-Tools/RM-Tools/wiki/RMsynth1D}}, which output an RM catalog in the standardised form \citep{vaneckRMTable2023PolSpectra2023Standards2023}. 

\subsubsection{Band 1 and Band 2}\label{sect:B1_B2}
While both Band 1 and Band 2 data were processed with the development version of the POSSUM pipeline \citep{purcell2017}, the Band 1 data were processed as single fields (individually for ``core'' and ``south'' fields), while the Band 2 data were first combined (interleaved for ``core'' and in the overlap region with the ``south'') and then split into smaller tiles ({$\sim$3.5~deg$^2$). The $I$$Q$$U$ spectra were extracted in the same manner from both the single fields and the smaller tiles, as described above. 

In order to create the most reliable $\RM$ catalog from the Band 1 and Band 2 data, we initially applied a cut in signal to noise of $\text{S/N}>6$ (in polarization) to Band 1 core data, and $\text{S/N}>8$ to Band 2 data\footnote{19 sources from Band 1 core out a final total number of 149 in the final catalog of the relevant Shapley region, actually survived with $6<\text{S/N}<8$. Since they did not fail any other of our quality metrics, they were kept. They all belong to Band 1 core, the most reliable dataset, and their median degree of polarization is $\sim 5\%$ which is $\sim 5$ times higher than the cut applied to this data, thus indicating they are not dominated by leakage. }.
In both Band 1 and Band 2 the angular resolution is $20''$, so for multiple sources whose separation is $<20''$, we only kept the one with a higher $\text{S/N}$, thus avoiding statistical bias from duplicated $\RM$ components. Band 1 was split into core and south data, so we merged both catalogs dealing with the overlap region between them by removing the south counterparts and keeping the core ones, due to a more reliable instrumental polarization correction in the latter. In Band 1 and Band 2 we only kept sources with Ifitstat<5 thus filtering out those that are too faint in total intensity or those with problems raised during the model fitting of Stokes $I$\footnote{See \href{https://github.com/CIRADA-Tools/RM-Tools}{https://github.com/CIRADA-Tools/RM-Tools} for more information.}. There were sources that are lying in the overlap region between Band 1 south and Band 2 observations. To deal with these, for those in Band 1 we kept sources with a distance to the nearest beam $<0.3\,\mathrm{deg}$, and distance to the tile $<1\,\mathrm{deg}$. This allowed us to remove sources potentially dominated by leakage and closer to the edges of the beams. For both Band 1 south and Band 2 sources we applied a cut in the degree of polarization $>1.5\%$ (Band 1 core sources were applied a $>1\%$ cut, due to a better widefield leakage correction in that field). Then, we only kept unique Band 2 sources, i.e., not detected in Band 1, since Band 2 has a smaller bandwidth in $\lambda^2$ thus higher uncertainty in the $\RM$s. 

\subsubsection{Combined final catalog}\label{sect:RMcatalog}
In the end, 369 sources from Band 2 and 2106 out of Band 1 data were kept\footnote{These 2475 sources are used in Sect. \ref{sect:grm} to estimate the foreground contribution from the Milky Way to the $\RM_{\text{obs}}$ values. }. Nonetheless, a further selection process was made to end up with a total 149 rotation measures. These where chosen to be those in the relevant region of the SSC that allows us to study its clusters and groups, comprising the combination of the on and off-target regions (see Fig. \ref{fig:subsamples}). Out of the final total 149 sources, only 12 of them are from Band 2 and no ``south'' sources made it to this final version.
The catalog has metrics to quantify the Faraday complexity of the sources, and we find that only 4\% of the on-target sources have hints of Faraday complexity. Furthermore, the data used in this work is unsuitable for detailed complexity analyses (eg.~$QU$-fitting) due to the sub-optimal instrumental polarization correction. 
Figure \ref{fig:StokesI_RMs} shows the Stokes $I$ image of ASKAP's fields with the positions of the final $\RM$s overlaid. The median error of the sources in the final $\RM$ grid is $\delta\RM = 1.98\,\mathrm{rad/m^2}$ and the density of sources is $36\,\RM/\mathrm{deg^2}$. The most relevant quantities for the polarized sources in the final catalog can be found summarized in Table \ref{tab:master_table} for an example set of ten of them.

\subsubsection{Embedded sources identification}
We searched for sources embedded in Shapley, i.e., within $\lesssim 16\,\mathrm{Mpc}$ of A3558 (at $z=0.048$). 4 sources were found out of the 46 sources in the detected sub-sample (see Section \ref{sect:sigma_RM_on_target} for the definition of it) to be embedded. Their component names are: J132928-313924, J133048-314323 (in the bridge sub-sample), J133503-313918 and J132803-314527 (in the clusters sub-sample).

\subsection{Thermal Sunyaev-Zeldovich map from Planck}\label{sect:planck}

When CMB photons propagate through a gas of hot thermal electrons (like a galaxy cluster) they are likely to scatter off them, and typically they gain energy from these collisions (inverse Compton scattering). The way the strength of this effect is parametrized is via the $y$-Compton parameter \citep{mroczkowskiAstrophysicsSpatiallySpectrally2019} defined as
\begin{equation}
    y = \frac{\sigma_T}{m_ec^2}\int P_e(l)\, dl,
\label{eq:ycompton}
\end{equation}
where again, the integral is along the line of sight, and $P_e=n_ek_BT_e$ is the pressure of the gas of electrons. $\sigma_T$ is the Thomson cross section. 

We have used the MILCA (Modified Internal Linear Combination Algorithm) full mission ComptonSZ map from the 2015 Planck results \citep{aghanimPlanck2015Results2016}. MILCA is a component separation algorithm that was developed to reliably map spatially localised emission as opposed to the diffuse emission associated with the CMB \citep{hurierMILCAModifiedInternal2013}. We downloaded a cutout of the Shapley Supercluster from Planck's Legacy Archive web page\footnote{\href{https://pla.esac.esa.int/\#maps}{Planck Legacy Archive}} with pixel size of $1.71'$. The relevant SSC region is $1.71\times 2.42\,\mathrm{deg}^2$ (see Fig. \ref{fig:subsamples}). 

There are other $y$-maps available in the literature that use the same initial Planck data and implement different algorithms for tSZ signal reconstruction, along with deprojection of the Cosmic Infrared Background (CIB) dust contribution \citep{mccarthyComponentseparatedCIBcleanedThermal2024}. The root-mean-squared noise of this map in the same SSC region is 4\% lower than for the MILCA Planck map. The median ratio of the map values from \citet{mccarthyComponentseparatedCIBcleanedThermal2024} and those from the MILCA Planck map is $1.1$, thus indicating minimal impact on results derived from the $y$-map. 

\section{Observational Results}\label{sect:obs_results}

In this section we use the POSSUM $\RM$s to study the statistical properties of the magnetized gas in the Shapley Supercluster Core (SSC), and how they vary as a function of distance from the cluster and group centers. 
 
\subsection{Galactic foreground subtraction and residual $\RM$s}\label{sect:grm}

\begin{figure*}[ht]
    \centering
    \includegraphics[width=0.95\linewidth,clip=true,trim=0cm 0cm 0cm 0cm]{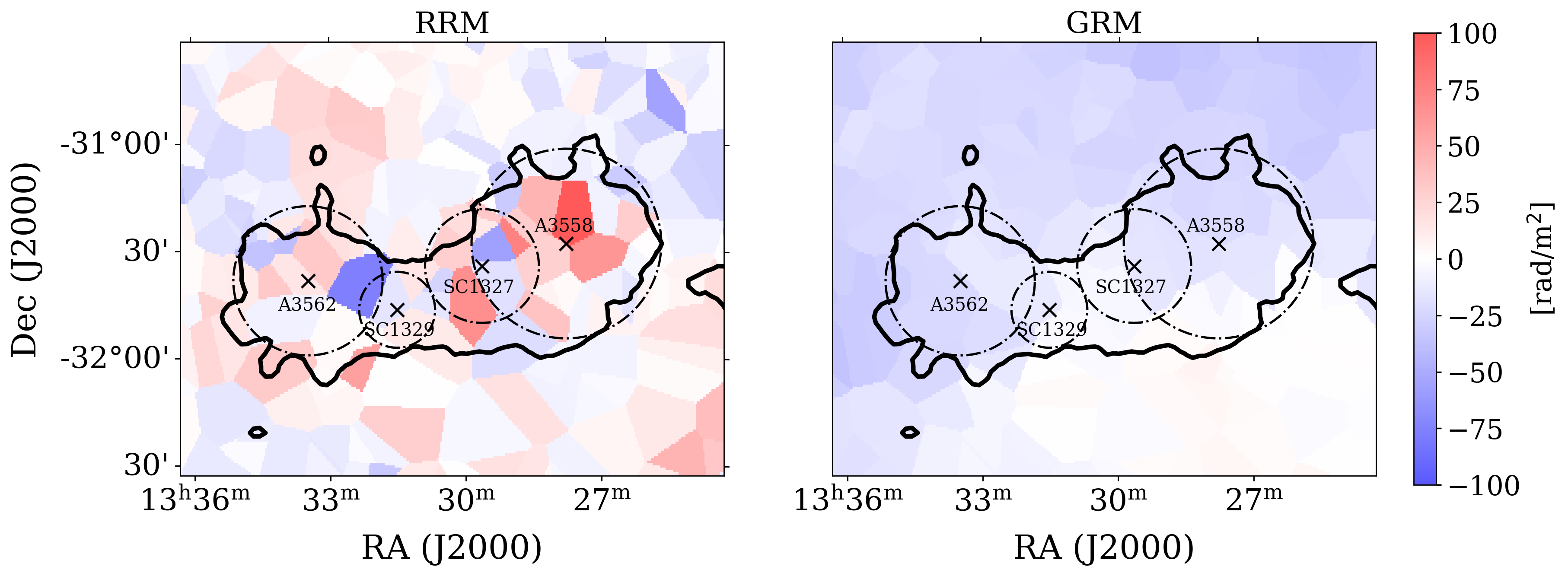}
    \caption{$\RRM$ and $\GRM$ maps made by interpolation to the nearest pixel of the estimated value at the particular position of a source from our catalog with the annulus method. The black contour corresponds to $y_{\text{bdry}}=4.24\times 10^{-6}$ (same as in Fig. \ref{fig:subsamples}, see Section \ref{sect:subsamples}). The centers of the clusters and groups are represented by the black crosses, while the dash-dotted circles represent their $r_{500}$. 
    Left:  Residual rotation measure map. The random nature of the patches in the map with similar sizes between them indicates that we have removed larger coherent RM structure of the foreground Galactic contribution, while retaining the information about the SSC.
    Right: $\GRM$ map. Opposite to the behavior of the patches in the $\RRM$ map, this map exhibits a large continuous gradient in the RMs, expected for the large scale Galactic contribution. }
    \label{fig:GRM_RRM_maps}
\end{figure*}

In order to study the $\RM$ properties related only to the SSC we need to remove correlated $\RM$ structures on scales larger than the SSC. The dominant contribution on such scales is expected to come from the Milky Way insterstellar medium (ISM), which we refer more generally to as the Galactic $\RM$ ($\GRM$), including the ISM, circum-galactic medium (CGM) and any magneto-ionized media associated with the Milky Way. We define the residual $\RM$ as 
\begin{equation}
    \RRM \equiv \RM_{\text{obs}} - \GRM.
\end{equation}
To estimate the $\GRM$ we will use the total 2475 POSSUM $\RM$s derived in Sect. \ref{sect:RMcatalog}. This grid has an areal density of $\rho_{\RM} \simeq41.5\,\mathrm{deg}^{-2}$. As a consequence, we did not employ commonly used methods such as the all-sky map of the Milky Way contribution \citep{hutschenreuterGalacticFaradayRotation2022}, which effectively loses information on scales $\lesssim 1\,\mathrm{deg}^{2}$. 
Instead, we have used the annulus method \citep{andersonProbingMagnetisedGas2024}, in which we can adapt the inner and outer annulus radii as relevant to the SSC system. 

\subsection{GRM from the annulus method}
In this method, for each $\RM$ value, we estimate the $\GRM$ at the source's position by defining an annulus and averaging\footnote{We did the same analysis with the median instead, with a resulting MAD standard deviation of the final $\RRM$ distribution consistent within the uncertainties with the one obtained using the mean.} over the $\RM$ values that lie inside it. The annulus is defined by two parameters: the inner exclusion radius $r_{\text{inner}}$ and the number of sources we want to average over. First we compute the distance from the source to every other source in the catalog and we keep the $50$ nearest ones whose distances are $>r_{\text{inner}}$. Thus, the furthest source defines the outer radius of the annulus. 

The parameters $r_{\text{inner}}$ and the number of sources to average over, were chosen to avoid removing the actual SSC $\RM$ structure, while accounting for the foreground Galactic contribution. The chosen $r_{\text{inner}}=1.7\, \mathrm{Mpc}$ ($0.5\,\mathrm{deg}$) has a size comparable to the average $r_{200}$ of the four objects in the Shapley core ($\Bar{r}_{200}=1.58\,\mathrm{Mpc}\lesssim r_{\text{inner}}$), thus we are not removing information inside the $r_{200}$ of the clusters and groups, while any contributions on larger scales are expected to be sub-dominant, as evidenced by cluster-derived magnetic field power spectra \citep{dominguez-fernandezDynamicalEvolutionMagnetic2019}. By choosing 50 sources to average inside the annulus, the average outer radius is $\Bar{r}_{\text{outer}}\approx 2.8 \,\mathrm{Mpc}\gtrsim 1.5 \,\Bar{r}_{200}$. The RRM errors are computed as the quadratic sum of the measurement errors ($\delta\RM$) and the standard error of the mean of the $50$ sources averaged over in the annulus for that particular source $i$
\begin{equation}
    \delta\RRM_i=\sqrt{(\delta\RM_i)^2 + \left(\sigma^{\text{ann,}i}_{\RM}\right)^2/N }.
    \label{eq:deltaRRM}
\end{equation}

In Fig. \ref{fig:GRM_RRM_maps} the RRM and GRM maps are shown. A smooth gradient over the SSC is seen in the $\GRM$, while the $\RRM$ shows a more irregular pattern. We compared our $\GRM$ results with the all-sky reconstruction of the Milky Way contribution in \citet{hutschenreuterGalacticFaradayRotation2022}. In contrast with the $\GRM$ gradient in Fig. \ref{fig:GRM_RRM_maps}, we found an almost constant $\GRM$ throughout the SSC compatible with the average value from the annulus method. This is consistent with the fact that this map loses information on scales lower than $\sim 1$ degree (due to the sparsity of the underlying input RM data). The median $\delta\GRM$ in the Hutchenreuter et al.~map is also $\sim2$ times larger than for the annulus method. We also checked for a potential correlation between the GRM and the dust polarization map \citep{ercegFaradayTomographyLoTSSDR22024} from Planck observations \citep{planckdust2020} finding no significant relation. This is supported by the fact that the thermal electrons, which are responsible for the GRM, do not necessarily reside in the dust-dominated region of the Galaxy. 

A comparison of the statistical properties of the $\RM_{\text{obs}}$ and the $\RRM$ is shown in Table \ref{tab:GRM}. The $\RRM$ distribution exhibits a mean and median closer to zero and a lower scatter than the observed $\RM$ distribution. Figure \ref{fig:RRM_GRM_hists} shows the histograms of the $\RM_{\text{obs}}$ and $\RRM$ distributions. In Appendix \ref{sect:grm_tests} we demonstrate the robustness of this method. 

\begin{figure}[ht]
    \centering
    \includegraphics[width=0.9\linewidth,clip=true,trim=0cm 1cm 0cm 3cm]{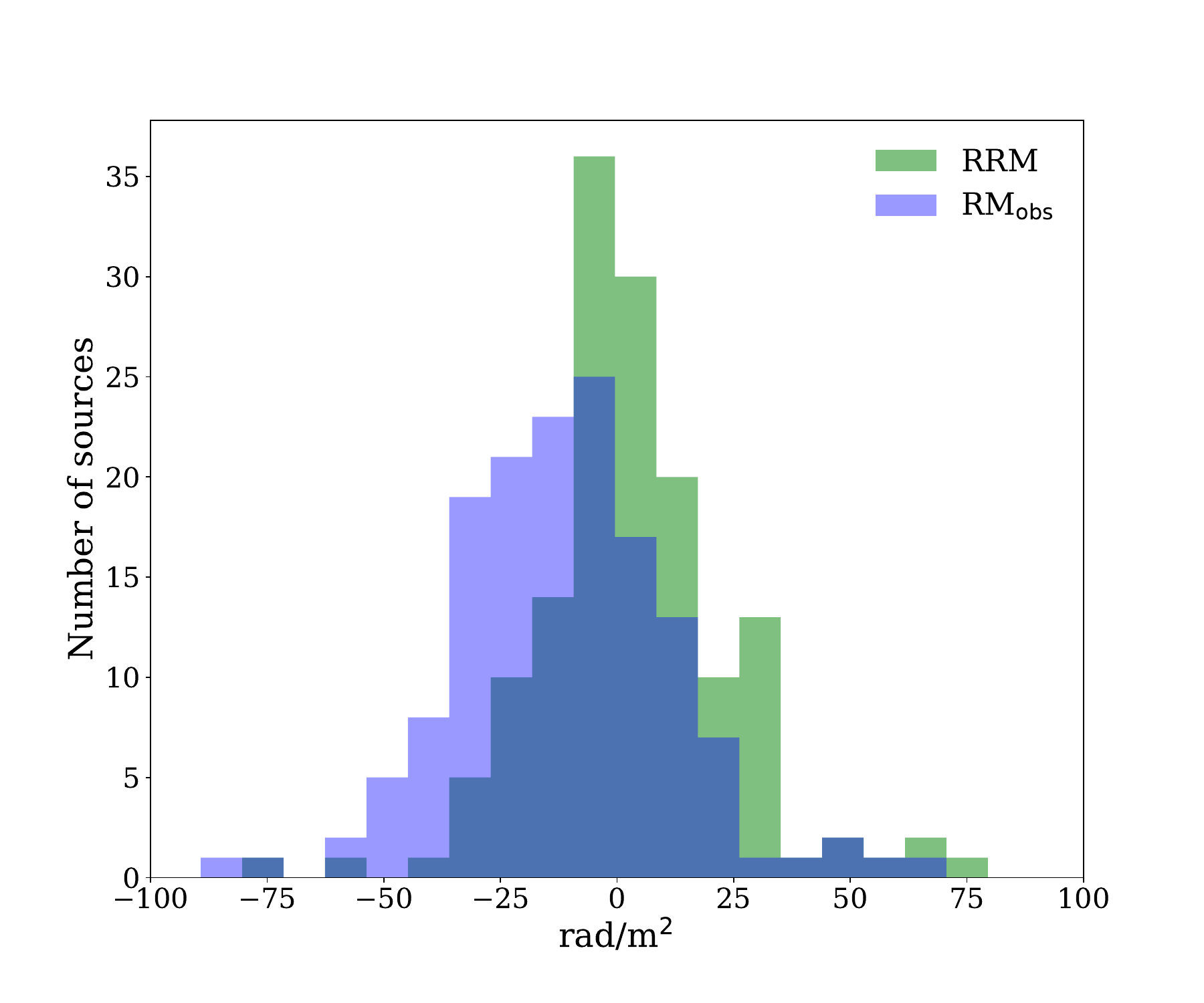}
    \caption{Histograms of observed rotation measures ($\RM_{\text{obs}}$) and residual rotation measures ($\RRM$) of the 149 sources used for our analysis. }
    \label{fig:RRM_GRM_hists}
\end{figure}

\begin{table}[h]
\caption{Mean, median and MAD standard deviation (Sect. \ref{sect:intro}) of the $\RM$ distributions before and after the removal of the Galactic foreground contribution with the annulus method.}
\centering
\renewcommand{\arraystretch}{1.1}
\begin{tabular}{c|ccc}
[rad/m$^2$] & Mean & Median  & $\sigma$   \\ \hline \hline
$\RM_{\text{obs}}$ & $-9.8\pm2.4$ & $-13.1\pm3.0$ & $21.9\pm1.6$ \\ 
$\RRM$ & $4.1\pm2.4$ & $0.9\pm2.9$ & $16.1\pm 1.4$
\end{tabular}
\tablefoot{The mean and median of the $\RRM$ are closer to the expected $\langle\RRM\rangle=0$ and the standard deviation $\sigma$ goes down by $\simeq 6\,\mathrm{rad/m^2}$.}
\label{tab:GRM}
\end{table}

\subsection{Definition of sub-samples}\label{sect:subsamples}

The 149 sources in the catalog we produced (see details in Sect. \ref{sect:askap}) sample the SSC, which corresponds to $\rho^{\text{SSC}}_{\RM}=36\,\RM$s/deg$^2$ ($ 3\,\RM$s/Mpc$^2$). 
We define the extent of the SSC region for study based on the Planck $y$-map, with the boundary between on-target and off-target regions defined as twice the $y$-map's rms noise level $y_{\text{bdry}}=2y_{\text{rms}}=4.24\times 10^{-6}$ (see Fig. \ref{fig:subsamples} for context), where 
\begin{itemize}
    \item Off-target $ \equiv y<y_{\text{bdry}}$,
    \item On-target $ \equiv y\geq y_{\text{bdry}}$.
\end{itemize}

The on-target region contains 46 sources. The off-target sub-sample contains 103 sources, which is a sufficiently large control sample that also avoids regions too far from the center. This avoids encountering contributions from other clusters and groups in the wider Shapley Supercluster area. The on-target region contains lines of sight through the Abell clusters and the SC groups. To study these contributions separately the on-target sample is further split into the clusters and bridge sub-samples. 

The bridge box (see Fig. \ref{fig:subsamples}) was defined to be parallel to the axis joining the centers of the Abell clusters. Despite the overlapping of the groups and clusters, this definition aims to isolate those lines of sight that mainly go through the SC groups (the bridge) inside their $r_{500}$, with 12 sources in this sub-sample. The remaining 34 sources that belong to the on-target region, are defined as the clusters sub-sample. These mainly represent lines of sight that cross the Abell clusters inside their $r_{500}$.

X-ray data from eROSITA could be used to trace thermal electrons and define the SSC boundaries. 
The main advantage of eROSITA compared to the XMM-Newton pointed observations is the larger and contiguous area coverage \citep{bulbulSRGEROSITAAllSky2024}. While eROSITA covers the whole of Shapley, given that the deeper XMM-Newton observations cover the entirety of the SSC, which is the region of interest for this work, no significant improvement is expected from eROSITA data in this context. Furthermore, the linear dependence of the tSZ effect with the electron density makes it a more suitable and sensitive tracer to define the SSC boundary, compared to the squared dependence of the X-ray emission.


\begin{figure}[ht]
    \centering
    \includegraphics[width=1\linewidth,clip=true,trim=0cm 3cm 0cm 6cm]{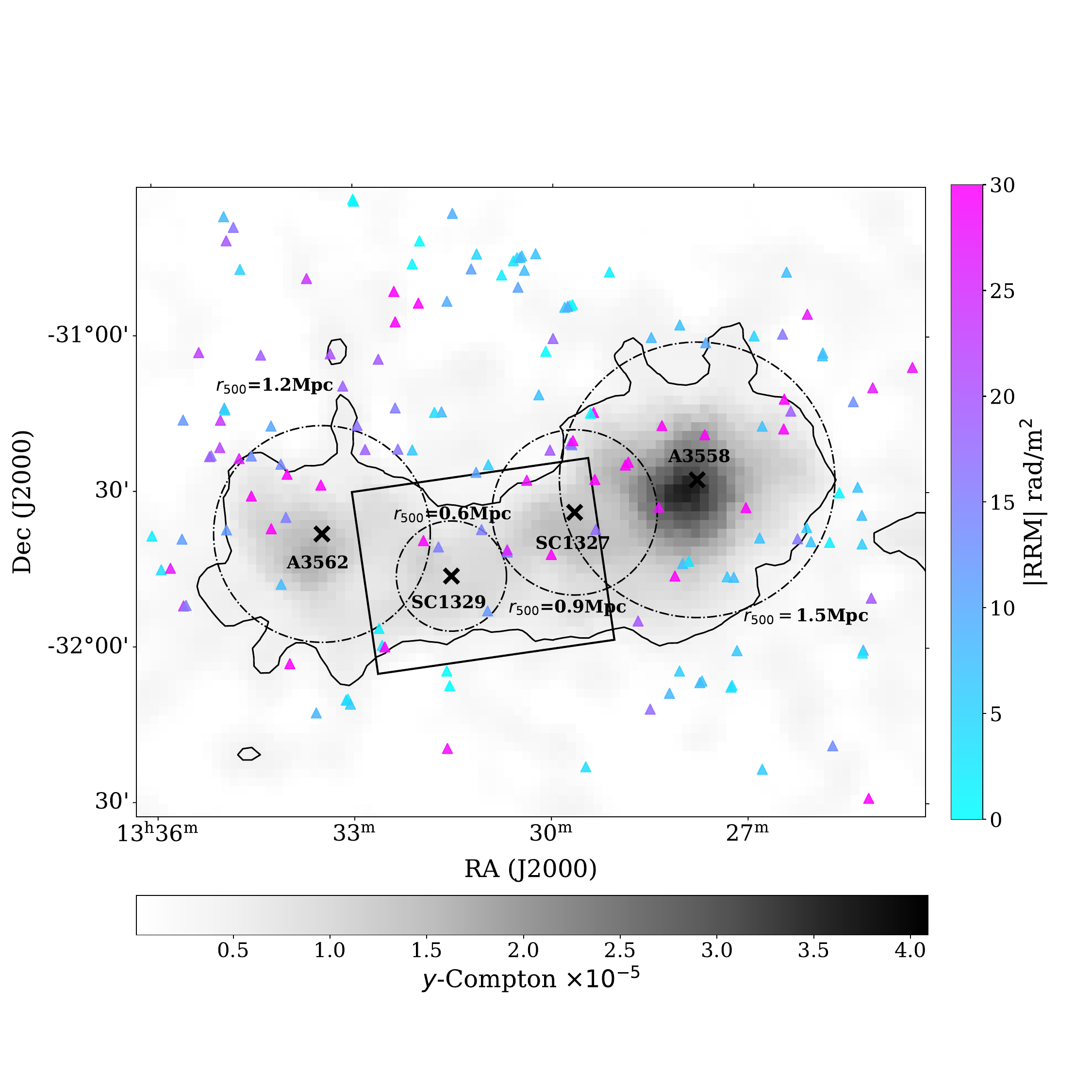}
    \caption{ tSZ Planck map of the A3558-A3562 clusters system as well as the two massive groups of galaxies SC 1327, SC1329. The triangles represent the locations of the background ASKAP radiogalaxies, and they are colored by their $\abs{\RRM}$ values. We have also represented the $r_{500}$ of the four objects as reference for the size of the system at the plane of the sky, along with their centers. The black contour represents the threshold we have used to define the boundary between the off-target and on-target regions: $y_{\text{bdry}}=2y_{\text{rms}}=4.24\times 10^{-6}$. The rectangular box defines the bridge. The sources inside it sample the region between the Abell clusters and outside their $r_{500}$, despite some overlapping effects. The counterpart on-target sources that lie outside the box, mainly sample the inside the of $r_{500}$ of the clusters, thus its name: the clusters sub-sample.}
    \label{fig:ShapS_y_RMs}
\end{figure}

\subsection{RM signature in the Shapley Supercluster Core}\label{sect:sigma_RM_on_target}

Hereafter, we will use ``$\RM$'' for the results derived from the $\RRM$s except where needed in case of a potential confusion. \\

The higher the scatter in a distribution of $\RM$s, the more likely it is that there is an excess contribution from magnetized gas. We calculate the excess $\RM$ scatter (MAD standard deviation, Sect. \ref{sect:intro}) of the SSC as
\begin{equation}
\small
    \sigma^{\text{SSC}}_{\RM} = \sqrt{\sigma^2_{\RM,\text{on-target}} - \sigma^2_{\RM,\text{off-target}}} = 30.52\pm 4.55 \,\, \mathrm{rad/m^2},
\label{eq:signal_obs}
\end{equation}
with a statistical significance of $6.7\sigma$. We found no significant variations in the exact value of the scatter after exploring minor variations in the value of $y_{\text{bdry}}$. 
The scatter in the on-target region is $\sigma_{\RM,\text{on-target}}=32.91\pm 4.18\,\, \mathrm{rad/m^2}$, while the off-target sample presents a scatter of $\sigma_{\RM,\text{off-target}}=12.32\pm 1.54\,\, \mathrm{rad/m^2}$. This value is higher than the expected $\sim6$-$8\,\, \mathrm{rad/m^2}$ for a random extragalactic background \citep{schnitzelerLatitudeDependenceRotation2010}. It implies that there is a contribution from magnetized gas outside the SSC, or that there remains uncorrected $\GRM$ contributions in the background $\RRM$s. However, $\sigma^{\text{SSC}}_{\RM}$ is not affected by this, since both sub-samples are equally affected by these contributions. 
Removing the 19 sources with $6 < \text{S/N} < 8$ leaves the statistics unchanged within the errors.


\subsection{RM signature within the clusters and bridge }\label{sect:obs_RM_scatter_bridge_clusters}

We investigate the $\RM$ sub-structure of the SSC by separately analysing the Abell clusters and the bridge region (where the SC two groups are).
Figure \ref{fig:ShapS_y_RMs} shows the $y$-map of the SSC, along with the $\abs{\RRM}$ of the sources and the bridge box. It also highlights the $r_{500}$ of the clusters and the groups.  
The excess $\RM$ scatter detected from these regions follows Eqn. \eqref{eq:signal_obs}, with the same off-target sample and replacing $\sigma_{\RM,\text{on-target}}$ with the RM scatter of the bridge and clusters sub-samples (Table \ref{tab:observed_sigma_RMs}). 

\begin{table}[ht!]
\caption{Detected excess in the $\RM$ scatter from the SCC, bridge and clusters sub-samples, along with their significance and the total number of sources inside each. }
\centering
\renewcommand{\arraystretch}{1.1}
\begin{tabular}{c|ccc}
Sample & Sources & $\sigma^{\text{Sample}}_{\RM}$ [rad/m$^2$] & Significance \vspace{1pt} \\  
 \hline \hline
   SSC & 46 & $30.52\pm 4.55$ & 6.7$\sigma$ \\
   Bridge & 12 & $25.32\pm 8.48$ & 3.0$\sigma$  \\
   Clusters & 34 & $27.20\pm 4.97$ & 5.5$\sigma$ 
\end{tabular}
\tablefoot{The off-target sub-sample is common for all three cases with a scatter of $\sigma_{\text{off-target}}=12.32\pm 1.54\,\, \mathrm{rad/m^2}$.}
\label{tab:observed_sigma_RMs}
\end{table}

The excess RM scatter within the bridge implies that  $\sigma^{\text{SSC}}_{\RM}$ is not completely dominated by the RM of the clusters. Furthermore, the fact that $\sigma_{\RM}^{\text{Bridge}}$ is compatible within the uncertainties with $\sigma_{\RM}^{\text{Clusters}}$, implies that we are probing a very similar range of $\sigma_{\RM}\propto n_eB_{\parallel}\sqrt{\Lambda L}$ (Eqn. \eqref{eq:sigmaRM}) in the clusters and bridge regions. This is supported by the electron density range computed in Appendix \ref{sect:obs_n_e_degeneracy}. 
We have also computed the excess $\RM$ scatter from each of the clusters, and found $\sigma_{\RM}^{\text{A3558}}=30.16 \pm 6.51$ and $\sigma_{\RM}^{\text{A3562}}=27.69 \pm 8.73$ at $4.6\sigma$ and $3.2\sigma$ respectively. These, further suggest that the range of $n_eB_{\parallel}\sqrt{\Lambda L}$ values is similar in the entire SSC region and its substructures, i.e., clusters combined, bridge, and each cluster separately. 


\subsection{Observed RM scatter profile with distance to nearest cluster/group}\label{sect:obs_RM_scatter_prof}

To further study how the scatter in the $\RM$ varies with distance from the clusters and groups centers, we 
follow \citet{andersonProbingMagnetisedGas2024} and \citet{osingaProbingClusterMagnetism2025}, and define the observable for the $\RM$ scatter profile with distance between independent sources as
\begin{equation}
    \mathfrak{S}_{\RM}(d_{\text{nrst}})\equiv \sqrt{\sigma_{\RM}^2(d_{\text{nrst}}) - \frac{\sum_{i=1}^{N}\delta\RRM_i^2}{N-1} - \sigma^2_{\RM, \text{off-target}}},
    \label{eq:scatter_prof_formula}
\end{equation}
where we correct for the off-target contribution and for the $\RRM$ measurement errors (see Eqn. \eqref{eq:deltaRRM}).
The distance $d_{\text{nrst}}\equiv r_{\text{nrst}}/r_{500}$ is the (projected) distance from a given source to the nearest cluster/group normalised by its $r_{500}$. This definition deals with the fact that in the SSC, the complicated structure and geometry does not have spherical symmetry, as in single cluster studies. For this profile, we use the on-target region sub-sample (see Sect. \ref{sect:subsamples}, Fig. \ref{fig:subsamples} and Fig. \ref{fig:ShapS_y_RMs}). The range of distances of our clusters and groups we are able to probe with these data is $ 0.3 \lesssim d_{\text{nrst}} \lesssim 1.8$.

\begin{figure}[ht]
    \centering
    \includegraphics[width=1\linewidth,clip=true,trim=0cm 0.8cm 0cm 2cm]{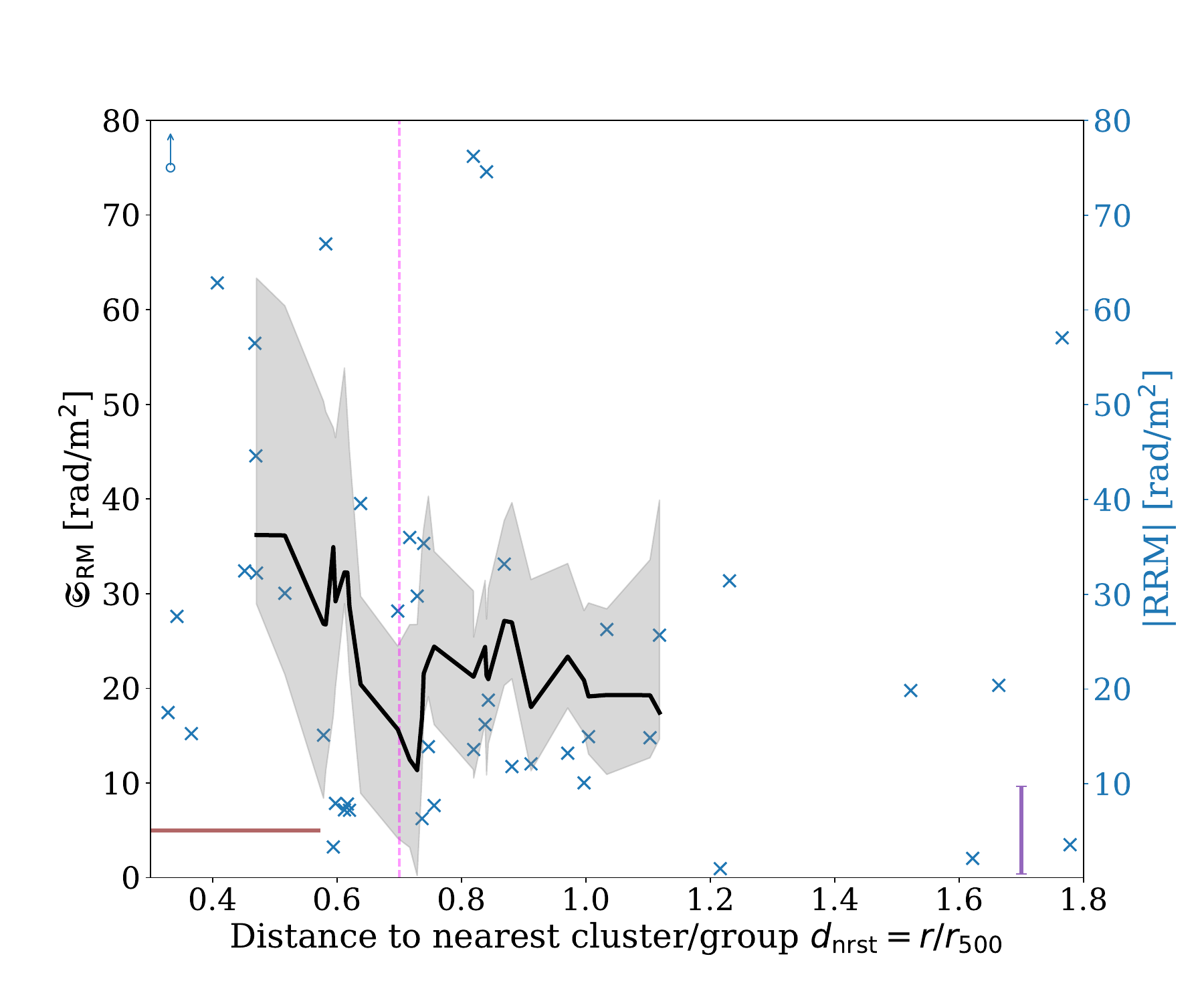}
    \caption{ Observed $\mathfrak{S}_{\RM}(d_{\text{nrst}})$ profile (black line). The gray shaded region is the 68$\%$ confidence level (C.L.) error region. The blue crosses represent the amplitude of the $\RRM$s of the on-target region, while the horizontal line is $\sigma_{\text{off-target}}$. The arrow-shaped data point corresponds to an outlier value of $\sim230$~rad/m$^2$. The vertical line is set at $0.7\,r_{500}$, which is the scale at which overlapping effects start to take place. The horizontal segment at the lower left corner represents the median window size of $\sim0.3\,r_{500}$. The vertical purple error bar (lower right corner) indicates the median on-target $\delta\RRM=4.6$rad/m$^2$. }
    \label{fig:RM_scatter_obs}
\end{figure}

To bin our data, we have used a sliding window approach, with fixed size of the window of $N=16$ data points, which translates to a median window size of $\sim0.3\,r_{500}$. Error bars have been obtained through bootstrap with $10^6$ resamples. The resulting behavior (Fig.~\ref{fig:RM_scatter_obs}) is statistically consistent with a moderate anti-correlation between the scatter in the $\RM$ and $d_{\text{nrst}}$, as a Spearman rank test yields a correlation coefficient of $-0.54$ with a p-value of $2\times 10^{-3}$. 
 
For $d_{\text{nrst}}>0.7\,r_{500}$ we find sources that are inside the $r_{500}$ of two different structures, i.e., a cluster and a group, thus we define this as the boundary at which overlapping structures begins. We quantified the trend with $d_{\text{nrst}}$ of sources unaffected by the overlapping of structures ($d_{\text{nrst}}\leq0.7\,r_{500}$) with a Spearman rank test with correlation coefficient of -0.65 and p-value of $0.03$. 
Beyond $>0.7\,r_{500}$ the profile appears to flatten (Fig.~\ref{fig:RM_scatter_obs}, right of the}vertical pink line). A Spearman rank test yields a correlation coefficient of 0.02, supporting this argument, but not with enough statistical significance (p-value of 0.9).
To assess the impact of the bridge and galaxy groups on the profile in Fig. \ref{fig:RM_scatter_obs}, we examined a sub-sample of sources, focusing on 19 sources whose l-o-s pass through the outer side of the cluster halos (counter to the bridge direction). After dividing the sample into two bins based on $d_{\text{nrst}}$, we find that the scatter still closely follows the original profile for the entire sample. Therefore, the inner structures of the bridge and groups do not significantly affect the profile. 

\subsection{Impact of small scale magnetic fields on RM statistics}\label{sect:beam_depol}

Magnetic fields coherent on scales smaller than our beam resolution, i.e., $\sim 20\,\mathrm{kpc}$, may play an important role contributing, for instance, to the total $\sigma_{\RM}^{\text{SSC}}$. They may also cause significant depolarization, such that within the SSC region we could be missing a population of polarized sources due to our selection criteria (Sect. \ref{sect:RMcatalog}). 
However, there is no clear evidence for a missing population 
given that the $\RM$ number densities are similar in the on-target and off-target regions ($\rho_{\RM}^{\text{off-target}} = 35\,\RM$s/deg$^2$ and $\rho_{\RM}^{\text{on-target}} = 37\,\RM$s/deg$^2$). 

To further assess the $\RM$ scatter due to small scale fields, which we denote by $\Sigma_{\RM}$ following the convention in \citet{osingaProbingClusterMagnetism2025}, we can use Burn's law \citep{burnDepolarizationDiscreteRadio1966, sokoloffDepolarizationFaradayEffects1998}
\begin{equation}
    p = p_0\,e^{-2\Sigma^2_{\RM}\lambda_0^4},
    \label{eq:burn_law}
\end{equation}
where $\lambda_0\simeq32\,\mathrm{cm}$ corresponds to the weighted mean of the observed  wavelength-squared channels of Band 1 observations ($\nu\simeq932\,\mathrm{MHz}$). 
We set $p_{\text{off-target}}\equiv p_0 = 5.3\%$, which is the median degree of polarization of the $\RM$ sources in the off-target sample. The median degree of polarization of the on-target sample is lower at $p_{\text{on-target}}\equiv p=4.0\%$, and therefore on average these sources suffer additional depolarization due to a $\Sigma_{\RM,\text{on/off}} = 3.6$ rad/m$^2$. 

This small-scale RM contribution could contribute to a decrease in $\mathfrak{S}_{\RM}$ at low values of $d_{\text{nrst}}$. The median degree of polarization of on-target sources inside and outside $0.7\,r_{500}$ is $p_{\text{on},\leq0.7}=3.7\%$ and $p_{\text{on},>0.7}=4.4\%$, respectively. Taking the ratio of these two, we find a depolarization of $\sim0.84$ between sources closer to the clusters' centers with respect to those outside. Using Eqn. \eqref{eq:burn_law}, we find $\Sigma_{\RM,\text{on-in/out}} = 2.9$ rad/m$^2$ (small in comparison to the total signal from SSC of $\simeq30$rad/m$^2$). These results indicate that beam depolarization appears to be a minor effect, likely playing a sub-dominant role in the overall $\RM$ statistics of our sample. Deeper observations and/or at higher frequencies, where depolarization effects are less, would be required to advance on this analysis. 
 
\section{Modeling results}\label{sect:model_results}

\subsection{Magnetic field strength estimates with single-scale models}\label{sect:single_scale}

To estimate the required magnetic field strength to explain the $\RM$ observations in the SSC region, we can use a single-scale model, e.g., \citet{gaenslerRadioPolarizationInner2001a}. This model assumes cells of fixed size $\Lambda$ and constant electron density $\Bar{n}_e$, where magnetic field strength is constant with random orientation with line of sight. The observed $\sigma_{\RM}$ and magnetic field strength are then related by 
\begin{equation}
    \sigma_{\RM} = \frac{e^3}{2\pi m_e^2c^4}\frac{\Bar{n}_eB}{2\sqrt{3}}\sqrt{L\,\Lambda},
\label{eq:sigmaRM}
\end{equation}
where $L=1.8\,\mathrm{Mpc}$ is the path length (for details on their estimation we refer to the reader to Appendix \ref{sect:L}). The coherence scale $\Lambda$ was fixed to $50$ kpc \citep{guidettiIntraclusterMagneticField2008, bonafedeComaClusterMagnetic2010, vaccaIntraclusterMagneticField2012, govoniSardiniaRadioTelescope2017}. The electron density estimates range from $\sim10^{-4}$ to $10^{-3}$ cm$^{-3}$ (Appendix \ref{sect:obs_n_e_degeneracy}, Table \ref{tab:B_single_scale}).
From Eqn. \eqref{eq:sigmaRM}, we determined the magnetic field strength
\begin{equation}
    B \simeq 1 \, \mu\mathrm{G} \, \left(\frac{\sigma_{\RM}\mathrm{[rad/m^2]}}{30}\right)\left(\frac{\Bar{n}_e[\mathrm{cm}^{-3}]}{4\times 10^{-4}}\right)^{-1}
\left(\frac{L\mathrm{[Mpc]}}{1.8}\frac{\Lambda\mathrm{[kpc]}}{50}\right)^{-1/2},
\label{eq:B_single_scale}
\end{equation}
in the bridge, clusters and SSC regions, ranging from $0.2\,\mu\mathrm{G}$ to $1.7\,\mu\mathrm{G}$ in the bridge and cluster regions for a wide range of electron density assumptions (see Appendix \ref{sect:single_scale_details}, Table \ref{tab:B_single_scale}).

\subsection{Gaussian random fields with MiRò}\label{sect:miro_results}

Here we investigate more realistic ICM models in order to obtain more robust magnetic field estimates, determine the field correlation with electron density (i.e.,~$\abs{\Vec{B}}\propto n_e^{\eta}$), and assess whether an additional gas component besides the clusters and groups in the SSC is required. 
Therefore, we use semi-analytic models for the ICM which go beyond constant density assumptions and single-scale magnetic field fluctuations by incorporating electron density profiles for clusters and power spectra motivated by magnetohydrodynamic (MHD) simulations onto a Gaussian random magnetic field model. In particular, we modeled the electron density for the clusters and groups with universal density profiles \citep{prattLinkingUniversalGas2022} (see \ref{sect:univ_profiles}). To implement multi-scale fluctuations in the magnetic field, we used the power spectrum from \citet{dominguez-fernandezDynamicalEvolutionMagnetic2019} that peaks at $230$ kpc.  
The MiRò code \citep{bonafedeMeasurementsSimulationFaraday2013, stuardiIntraclusterMagneticField2021} is capable of incorporating the aforementioned elements, thus we use MiRò to model the SSC. Additional details on how this is done, can be found in Appendix \ref{sect:miro_B_methods}. 

\subsubsection{RM signatures predicted by MiRò}\label{sect:miro_signals}

\begin{figure}[ht]
    \centering
    \includegraphics[width=1\linewidth,clip=true,trim=0cm 0.8cm 0cm 2.2cm]{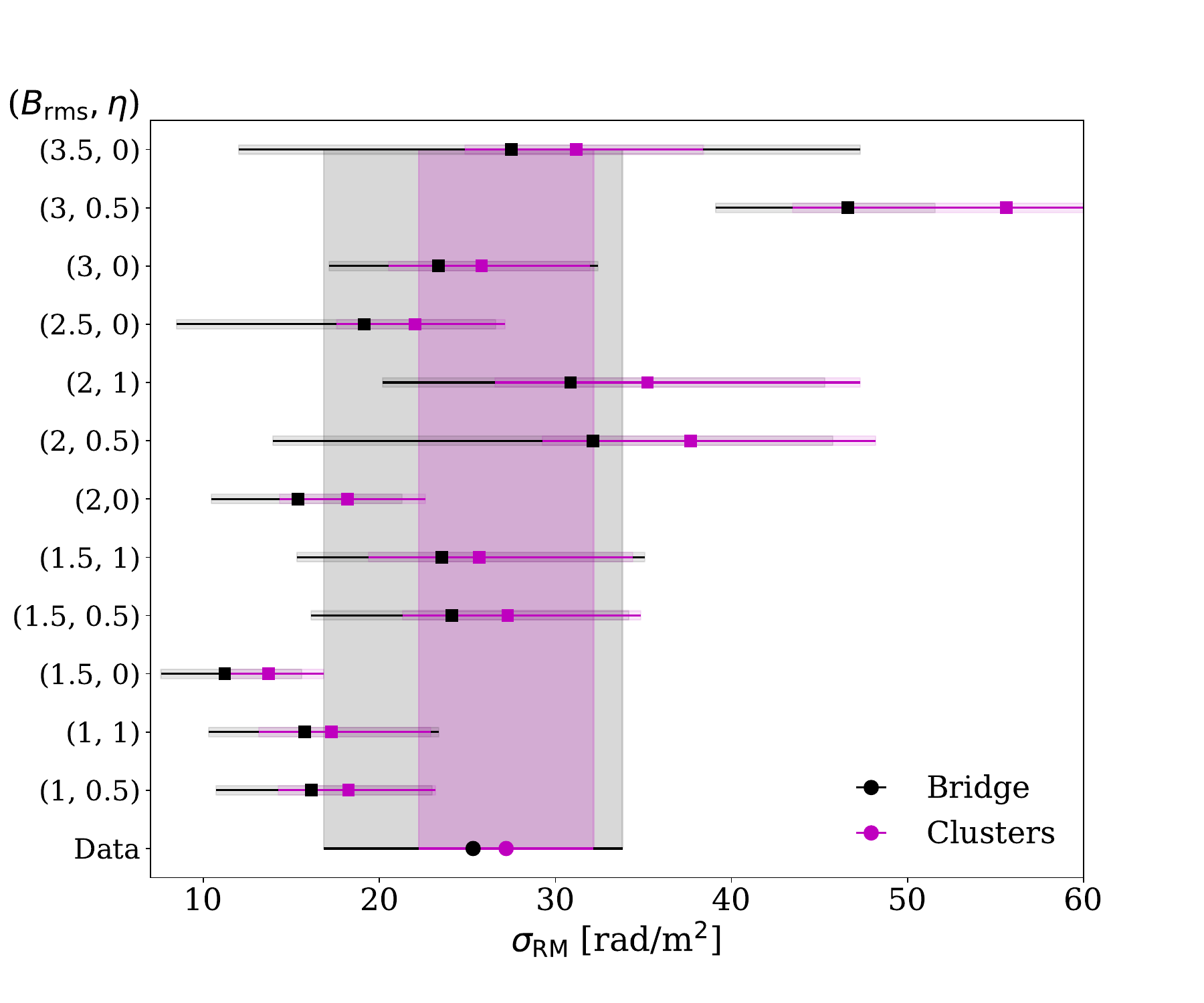}
    \caption{Visual comparison of the excess $\RM$ scatter in the bridge and clusters regions between the observations (bottom line) and the Gaussian Random field models from MiRò (computed following Appendix \ref{sect:miro_RM_scatter_method}). Horizontally, we represent the 68$\%$ C.L. error region for each model. Vertically we outline the statistical error estimated for the observations for visual comparison reasons. $B_{\text{rms}}$ is in $\,\mu\mathrm{G}$.}
    \label{fig:miro_obs_signal_comparison}
\end{figure}

We compute the excess $\RM$ scatter from the bridge and clusters regions for the MiRò models in the following parameter space: $B_{\text{rms}}= 1\,\mu\mathrm{G}-3.5\,\mu\mathrm{G}$ , $\eta=0,\,0.5,\,1$, where $\eta$ parametrizes the correlation between magnetic field strength and electron density $\abs{\Vec{B}}\propto n_e^{\eta}$.
Following a Monte-Carlo and bootstrap method (see Appendix \ref{sect:miro_RM_scatter_method}) we obtained $\sigma_{\RM}^{\text{Bridge}}$, $\sigma_{\RM}^{\text{Clusters}}$ shown in Fig. \ref{fig:miro_obs_signal_comparison} in comparison to the observed values. The bottom line shows the observed excess $\RM$ scatter in the bridge and clusters with its error bars, and vertically we show the estimates from the MiRò models.

 With this statistic only, despite the fact that there is a significant amount of overlap between models due to the large associated uncertainties, we can already rule out some models. The criterion to decide was whether the predicted $\sigma_{\RM}^{\text{Clusters}}$ is compatible, within the uncertainties (at the 1$\sigma$ level), with the observed one. Given that the power spectrum is derived from cluster simulations and there is no evidence that indicates it should also hold in the bridge region, MiRò is expected to perform best at the clusters. Table \ref{tab:signals_miro} shows the predicted $\RM$ scatter only of these models. 

Regarding the bridge region, note that only the inclusion of the two massive groups in the modeling is enough to explain the observations within the uncertainties for most models. Therefore, the contributions to the observed excess $\RM$ scatter from the clusters and groups dominates over other magnetized gas in the SSC. Hence, at the level of our data, there is no motivation to add an additional RM component (eg. from filamentary gas) in the intercluster region.

\subsubsection{Comparison between modeled and observed $\mathfrak{S}_{\RM}(d_{\text{nrst}})$}\label{sect:scatter_profs_miro_obs_comparison}

To better discriminate between models, we have computed scatter profiles of the $\RM$ with distance to the nearest cluster/group to compare the predictions from MiRò with the results in Sect. \ref{sect:obs_RM_scatter_prof}. To calculate the profiles from the mock $\RM$ maps, we binned the region with a sliding window of fixed step size $\Delta d_{\text{nrst}}=10^{-2}$, and the $68\%$ C.L. error region was computed through bootstrap with $10^4$ resamples. For more details see Appendix \ref{sect:methods_miro_profiles}. In Figure \ref{fig:scatter_profiles} we plot the resulting profiles as well as the observed profile from Sect. \ref{sect:obs_RM_scatter_prof}.

\begin{figure}[ht!]
    \centering
    \includegraphics[width=1\linewidth,clip=true,trim=0cm 0.8cm 0cm 2cm]{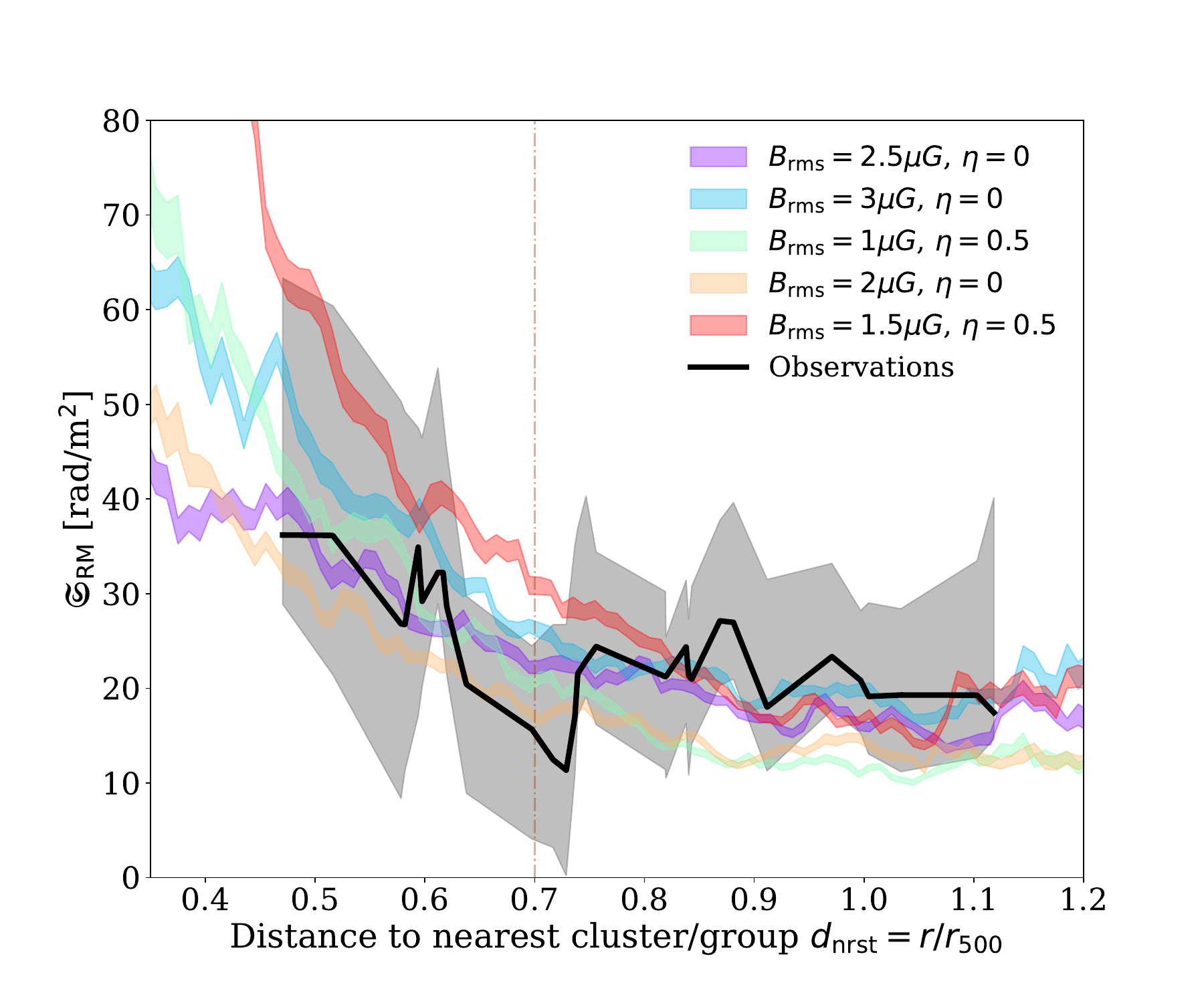}
    \caption{$\mathfrak{S}_{\RM}$ profiles. The color-shaded regions represent the $68\%$ C.L. error regions for the MiRò profiles (made with a sliding window of fixed size $0.01\, r_{500}$). We show the five MiRò models with highest Bayesian evidence. The black line represents the observed scatter profile and the gray shaded region represents its $68\%$ C.L. error region. The vertical line is at $0.7\,r_{500}$, where overlapping structures start to dominate. 
    }
    \label{fig:scatter_profiles}
\end{figure}

To determine which model best reproduces the observed profile we defined a Gaussian likelihood function $\mathcal{L}_k$ for each MiRò model $k$ as follows:
\begin{equation}
\small{
    \mathcal{L}_k = \frac{1}{\sqrt{(2\pi)^N\Pi_{i=1}^{N}\sigma^2_{\text{tot},i,k}}}\exp{\left(-\frac{1}{2}\sum_{i=1}^N\frac{(\mathfrak{S}_{\RM,i}^{\text{MiRò},k}-\mathfrak{S}_{\RM,i}^{\text{Obs}})^2}{\sigma^2_{\text{tot},i,k}}\right)}\, ,
    \label{eq:likelihood}
}
\end{equation}
where $\sigma^2_{\text{tot},i,k} = (\Delta\mathfrak{S}_{\RM,i}^{\text{MiRò},k})^2+(\Delta\mathfrak{S}_{\RM,i}^{\text{Obs}})^2$ is the total uncertainty in $\mathfrak{S}_{\RM}$ from both the model and the data, as computed through bootstrapping. 
We computed the Bayesian evidence for each model which we can simply identify with the likelihood\footnote{This holds due to the fact that we have a fixed parameter space (see Appendix \ref{sect:methods_miro_profiles}).}. Among the models tested, the one that provides the best description of the observational data is the one with the largest value of $\ln{\mathcal{L}_k}$. To compare the performance between different models we compute Bayes factors $\mathcal{B}_{jk}\equiv\mathcal{L}_j/\mathcal{L}_k$ and use the \citet{kassBayesFactors1995} scale (K\&R). For more details on the Bayesian model selection analysis see Appendix \ref{sect:methods_miro_profiles}.

To do this comparison, we used the value of the $\RM$ maps with a $\sigma^{\text{Clusters}}_{\RM}$ compatible with the observed excess $\RM$ scatter in the clusters (see Fig. \ref{fig:miro_obs_signal_comparison}) only at the location of the on-target $\RM$s and applied Eqn. \eqref{eq:scatter_prof_formula} with the same sliding window parameters applied to the observational profile. Even though MiRò is modeling overlapping structures, it is not accounting for the likely complex interactions between them. Therefore, we only compared the models to the data in the bins satisfying $d_{\text{nrst}}\leq 0.7$, where the overlapping structures are not dominating the behavior of the profiles. For values $>0.7\,r_{500}$ all profiles exhibit a similar flattening and most of them lie within the uncertainties of the observed profile. Thus, in order not to bias our conclusions from the Bayesian model selection with the performance of the profiles in that range of distances, we do not include them in this analysis. 

\begin{table}[h]
    \caption{ Bayesian evidence, Bayes factors and K\&R scale result for the MiRò model selection analysis. }
    \centering
    \renewcommand{\arraystretch}{1.1}
    \begin{tabular}{c|ccccc}
        ($B_{\text{rms}},\,\eta$) & (2.5, 0) & ($3$, $0$) & ($1$, $0.5$) & ($2$, $0$) & ($1.5$, $0.5$) \\
        \hline \hline
        $\ln{\mathcal{L}_k}$ & -43.6 & -44.2 & -45.9 & -46.1 & -48.3 \\
        $\ln{\mathcal{B}}$ & ... & 0.6 & 2.3 & 2.4 & 4.6 \\
        K\&R & ... & Barely & Positive & Positive & Strong
    \end{tabular}
\tablefoot{We compared the value of the $\mathfrak{S}^{\text{MiRò}}_{\RM}(d_{\text{nrst}})$ profiles at the position of the on-target sources in the range $\leq 0.7\,r_{500}$. More details can be found in Appendix \ref{sect:methods_miro_profiles}. $B_{\text{rms}}$ is in $\,\mu\mathrm{G}$. }
\label{tab:bayes_miro}
\end{table}

In Table \ref{tab:bayes_miro} we show the results of the Bayesian model selection analysis. The model with a highest Bayesian evidence corresponds to the model with the best predictive performance: $B_{\text{rms}}=2.5\,\mu\mathrm{G}$, $\eta=0$. We show the Bayes factors of all other models with respect to it sorted by their performance up to the first one with strong evidence against it. We find barely worth mentioning evidence against the $B_{\text{rms}}=3\,\mu\mathrm{G}$, $\eta=0$ model, and positive evidence against the best performing $\eta=0.5$ model.

Another approach is to use the scatter of the $\RM$ as a function of electron density, $\mathfrak{S}_{\RM}(n_e)$. This quantity could, in principle, provide a more direct mapping of the profile of the $\RM$ scatter from the center of the cluster towards its outskirts. However, the overlapping of structures in the SSC geometry, gives $\mathfrak{S}_{\RM}(d_{\text{nrst}})$ the advantage of clearly showing at what scale these become important. Moreover, to estimate the electron density from the observations some assumptions need to be made, such as constant density and temperature over the path length (see Appendix \ref{sect:obs_n_e_degeneracy}). 
Nonetheless, we show the profiles with respect to electron density and the comparison between models and data with the same Bayesian model selection analysis in Appendix \ref{sect:RM_scatter_profiles_n_e}. From these profiles the data favors the model with $B_{\text{rms}}=2.5\,\mu\mathrm{G},\,\eta=0$, which is in agreement with the above results from the comparison with the profiles with $d_{\text{nrst}}$. The five models with highest Bayesian evidence are the same as those from the profiles with $d_{\text{nrst}}$. However in this case, the second best performing model is  $B_{\text{rms}}=1\,\mu\mathrm{G}$, $\eta=0.5$, with barely worth mentioning evidence against it.

\subsubsection{Estimated magnetic field strength from MiRò}\label{sect:miro_B_avg}

The average magnetic field strength can be estimated in the bridge and clusters regions from the magnetic field 3D cubes generated by MiRò, but retains significant systematic uncertainty due to the model selection ambiguity. Applying Eqn. \eqref{eq:B_avg} to the model with the best performance according to the results from Section \ref{sect:scatter_profs_miro_obs_comparison}, i.e., $B_{\text{rms}}=2.5\,\mu\mathrm{G}$, $\eta=0$ we obtained $\langle \abs{\Vec{B}} \rangle\simeq 1.7\,\mu\mathrm{G}$ both in the bridge and clusters regions with a statistical error of $\sim 0.1\,\mu$G.

\subsection{Comparison to cosmological MHD simulations: SLOW}\label{sect:slow_results}

MiRò provides a significant improvement in the ICM and magnetic field modeling in clusters with respect to single-scale models. Nevertheless, it lacks a realistic description of the complex interactions between the clusters and groups in the SSC (it only assumes non-interacting spheres with a given density profile), or variable magnetic field coherence lengths (the same power spectrum was assumed to hold at all scales and distances in the entire SSC). However, cosmological MHD simluations provide a solution to these problems.

Therefore, we extend our analysis to compare the RM data from Section \ref{sect:obs_results} to modeling obtained from the SLOW (Simulating the Local Web)\footnote{\url{https://www.usm.lmu.de/~dolag/Simulations/\#SLOW}} simulation set \citep[][]{Dolag2023, bossSimulatingLOcalWeb2024}. SLOW is a cosmological MHD simulation constrained to match the local Universe large scale structure.
We refer the reader to Appendix \ref{sect:slow_methods} for the simulation setup details. 
From the simulation, we identified analogs of the SSC groups and clusters with the following properties \citep[see discussion in][]{Seidel2024}:
\begin{itemize}
    \item SLOW-A3562: $M_{500} = 5.3\times 10^{14} M_\odot$, $r_{500} = 1.2$ Mpc
    \item SLOW-A3558: $M_{500} = 1.46\times 10^{15} M_\odot$, $r_{500} = 1.7$ Mpc
    \item SLOW-SC1329: $M_{500} = 2.25\times 10^{14} M_\odot$, $r_{500} = 0.91$ Mpc
    \item SLOW-SC1327: $M_{500} = 2.04\times 10^{14} M_\odot$, $r_{500} = 0.88$ Mpc
\end{itemize}
The radii were obtained from the $M_{500}$ following the usual self-similarity relation $M_{500}\propto r_{500}^3$. The masses from the analog clusters and groups differ from the observed ones by a factor ranging from $0.02$ (SC 1327) to $3.5$ (SC 1329) larger. The $r_{500}$ is underpredicted by a factor $0.02$ for SC 1327 and overpredicted by a factor $0.1$-$0.5$ for A3558 and SC 1329, respectively, while SLOW-A3562 has a radius equal to the observed one. 

Within the simulation adiabatic compression and dynamo processes amplify a seed magnetic field. This simulated magnetic field ($B_{\text{sim}}$) produces realistic values in galaxy clusters but the radial decline is too steep due to resolution effects \citep[see][]{Steinwandel2022_dynamo}, hence filament magnetic fields are under-estimated.
To study the RM in filamentary regions between the clusters, several magnetic field models are applied to rescale the simulated magnetic field.
Namely: a constant plasma-$\beta$ ratio between magnetic and thermal pressures ($B_{\beta}$); turbulent and magnetic pressure equilibrium ($B_{\mathcal{F}}$); a magnetic flux conservation during amplification by 3D compression mechanism ($B_{\text{ff}}$); and two turbulent dynamo scenarios, where one of them operates for a range of densities $n_e \gtrsim 10^{-4}$ cm$^{-3}$ ($B_{\text{dyn},\downarrow}$) with an attached decline at $n_e < 10^{-4}$ cm$^{-3}$ to match observations for filament magnetic fields \citep{carrettiMagneticFieldEvolution2023}, and the other is a pure saturated dynamo ($B_{\text{dyn},\uparrow}$), see \citet{bossSimulatingLOcalWeb2024} for details. 

To understand how these models predict different correlations between magnetic field strength and electron density (in analogy to the $\eta$ parameter in the modeling with MiRò), we have fit these two quantities in the electron density range $4\times 10^{-5} \leq n_{e} \, \text{\small [cm$^{-3}$] } \leq 8\times 10^{-4}$, where the comparison between the SLOW models and the data is performed (see Appendix \ref{sect:slow_method_RM_scatter}). We found that all amplification scenarios are well fitted by a power law with: $\eta_{\text{sim}}=0.44$, $\eta_{\beta}=0.53$, $\eta_{\mathcal{F}}=0.50$, $\eta_{\text{ff}}=0.67$, $\eta_{\text{dyn},\downarrow}=0.51$ and $\eta_{\text{dyn},\uparrow}=0.50$. 

\begin{figure*}
    \centering
    \includegraphics[width=0.9\linewidth]{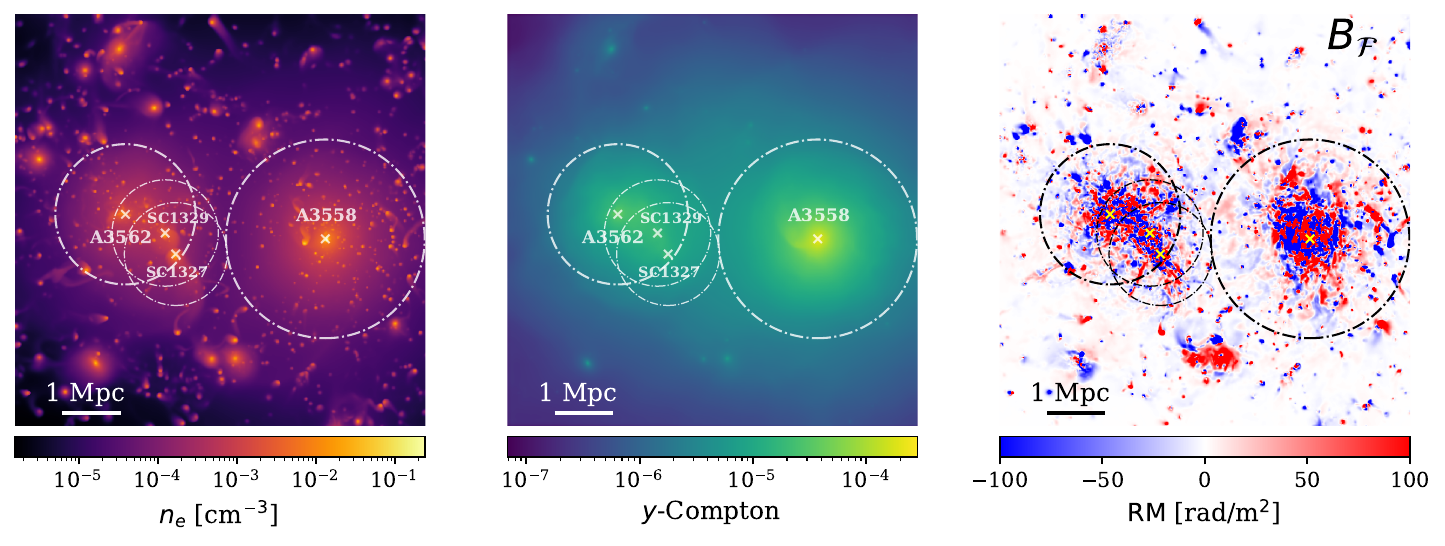}
    \caption{ Zoom-in simulation of the Shapley analog found in the SLOW simulation. From left to right: electron number density, $y$-Compton parameter, $\RM$ map for the dynamo $\downarrow$ amplification mechanism. We show the $r_{500}$ of the two clusters and the two groups, namely: $1.7, 1.2, 0.91$ and $0.88$ Mpc in decreasing order of mass.}
    \label{fig:slow_shapley}
\end{figure*}

Using SLOW it is possible to obtain maps of different physical effects and observables, e.g., X-ray surface brightness, tSZ effect, temperature, electron density. As an example, Fig. \ref{fig:slow_shapley} shows the electron number density and $y$-Compton maps, along with the $\RM$ map for one of the dynamo mechanisms of the Shapley analog found in the SLOW simulation. While generally well reproducing the clusters and groups' masses and sizes, it is worth mentioning that there are two main differences with respect to the observations: the projected distance between the Abell clusters is a factor $0.8$ lower, and the positions of the groups do not exactly match the observations. 

Using mass weighted magnetic field maps for the different amplification mechanisms, we have computed the average magnetic field in the bridge and clusters regions following Eqn. \eqref{eq:B_avg}\footnote{However, working with a 2D map instead of a 3D cube, implies we only need to sum over the $N_{\Omega}$ pixels. Thus, $N_z=1$ and we do not perform any sum over the line of sight. }. All models provide magnetic field strengths of the order of $\sim 1$ to 3 $\,\mu\mathrm{G}$, both in the bridge and clusters regions, except the $B_{\text{sim}}$ model that yields $\sim 0.1\,\mu\mathrm{G}$. The statistical errors are $\lesssim 10^{-3}\,\mu\mathrm{G}$.

\subsubsection{RM signatures predicted by SLOW}\label{sect:RM_scatter_slow}

We determine $\sigma_{\RM, \text{SLOW}}^{\text{Bridge}}$ and $\sigma_{\RM, \text{SLOW}}^{\text{Clusters}}$ from the SLOW SSC analog in a similar manner to the observations (see \ref{sect:slow_method_RM_scatter}). We have estimated the excess $\RM$ scatter in these regions using the $\RM$ maps derived from all SLOW magnetic field amplification mechanisms. 

\begin{figure}[ht!]
    \centering
    \includegraphics[width=1\linewidth,clip=true,trim=0cm 1cm 0cm 3cm]{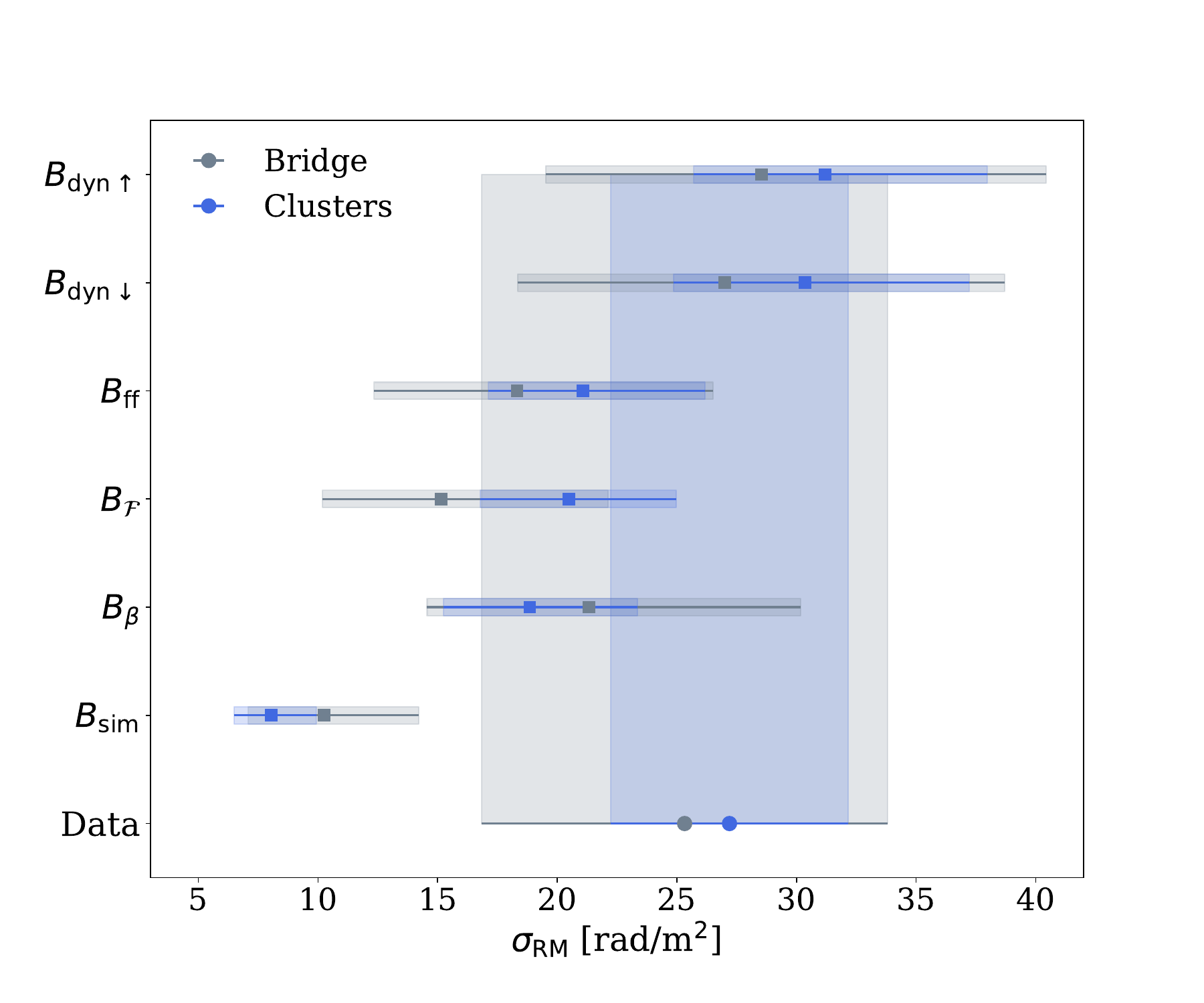}
    \caption{Visual comparison of the excess $\RM$ scatter in the bridge and clusters regions between the observations (bottom line) and all magnetic field amplification mechanisms implemented in SLOW. Horizontally, we represent the 68$\%$ C.L. error region for each model. Vertically we outline the statistical error estimated for the observations for visual comparison reasons. }
    \label{fig:slow_obs_signal_comparison}
\end{figure}
Based on $\sigma_{\RM, \text{SLOW}}^{\text{Clusters}}$ we find that the $B_{\text{sim}}$ model is inconsistent with observations at $4.6\sigma$, and therefore can be ruled out (see Fig. \ref{fig:slow_obs_signal_comparison}). This implies that we can rule out the possibility that in the gas between the Abell clusters no further amplification of the fields is taking place. 

All other mechanisms are able to reproduce the observed scatter in the clusters. The $B_{\beta}$ model, while showing a larger $\RM$ scatter in the bridge than in the clusters region, it also predicts a $\sigma_{\RM}^{\text{Clusters}}$ that is consistent with observations at the $1\sigma$ level. Moreover, similarly to what we found from the Gaussian random field modeling with MiRò, with this statistic we find overlap between these models within the uncertainties, thus being unable to discriminate between them. Table \ref{tab:slow_scatters} lists the values of the excess $\RM$ scatter shown in Fig. \ref{fig:slow_obs_signal_comparison}.

\subsubsection{Comparison between simulated and observed $\mathfrak{S}_{\RM}(d_{\text{nrst}})$}\label{sect:slow_scatter_profiles}

In order to better discriminate between models, using the $\RM$ maps derived from the simulation, we explore how the $\RM$ scatter behaves with distance to the center of the clusters for all amplification mechanisms considered. Details of how these profiles were constructed are provided in \ref{sect:slow_method_scatter_profiles}.
We cannot directly compare with the observational results in Sect. \ref{sect:obs_RM_scatter_prof}, because of the differences in projected distance between the clusters and the location of the groups. Therefore, we defined $d_{\text{nrst}}$ this time as the distance to the nearest cluster only, without including the groups. We computed a profile of the $\RM$ scatter from the observations with this new definition of distance, using Eqn. \eqref{eq:scatter_prof_formula}. 

\begin{figure}[ht!]
    \centering
    \includegraphics[width=1\linewidth,clip=true,trim=0cm 1cm 0cm 2.6cm]{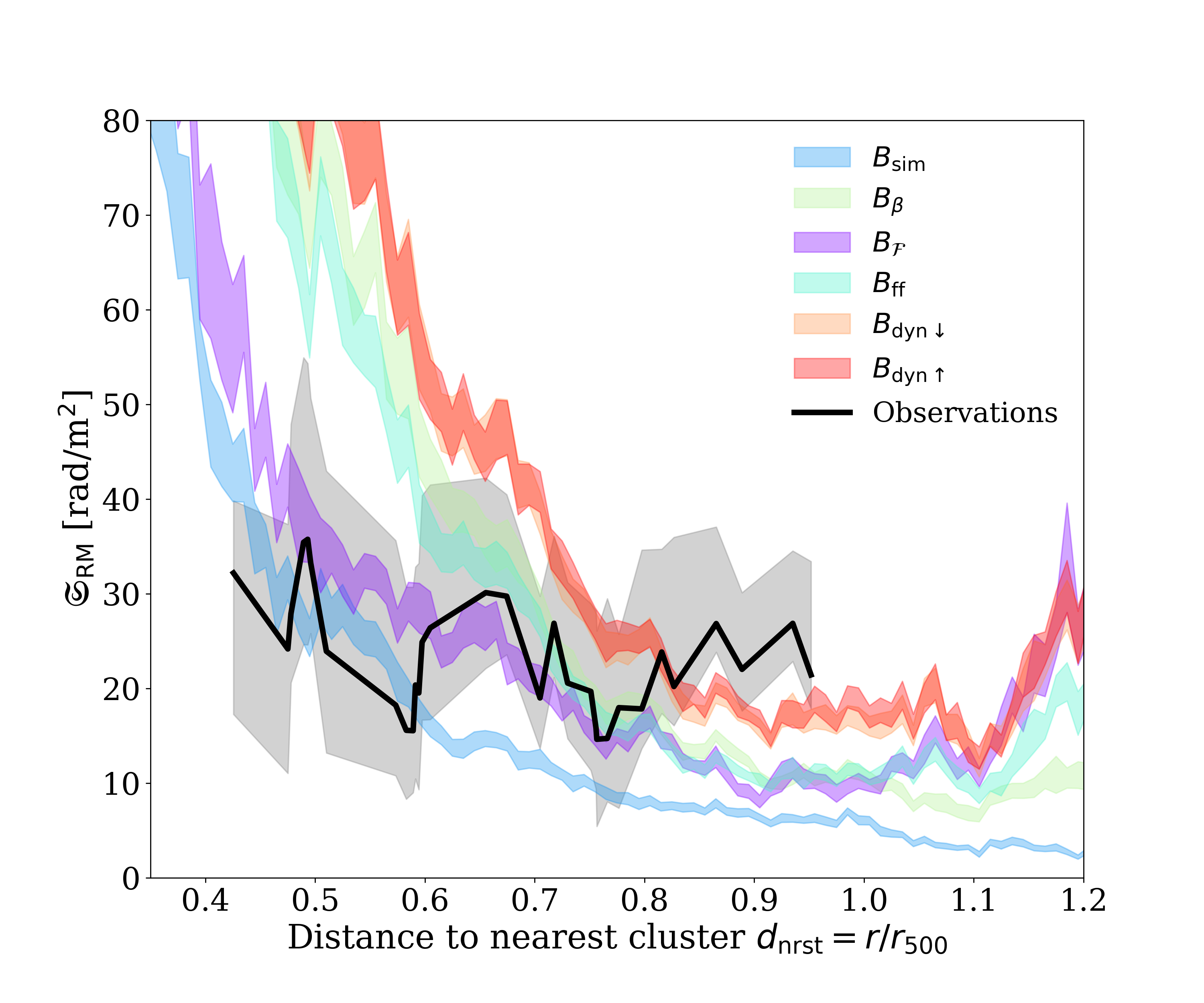}
    \caption{ $\mathfrak{S}_{\RM}$ profiles with distance to the nearest cluster only. The black line represents the observed scatter profile, computed using a fixed sliding window of $N=16$ points. The gray shaded region represents its $68\%$ confidence intervals obtained with bootstrapping resampling. The observational $d_{\text{nrst}}$ values have been rescaled by a factor $\times 0.8$ to account for the difference in projected distance between the SLOW analog clusters and the real A3558-A3562. 
    The colored regions are the SLOW $68\%$ C.L. error regions (made with a sliding window of fixed size $0.01\, r_{500}$) for the $\RM$ maps corresponding to all amplification mechanisms of the magnetic field in the intercluster region.
    }
    \label{fig:scatter_profiles_SLOW}
\end{figure}

Figure \ref{fig:scatter_profiles_SLOW} shows the new observational $\mathfrak{S}_{\RM}(d_{\text{nrst}})$ profile and the equivalent SLOW profiles of all amplification mechanisms. We have performed a Spearman rank test to the observed profile to asses its flatness, obtaining a correlation coefficient of $-0.36$ with a p-value of $0.05$. This indicates that no statistically significant anti-correlation is found, in contrast with the result from Section \ref{sect:obs_RM_scatter_prof} when we included the groups into the analysis. 

To determine the performance of these models we implemented a Bayesian analysis similar to that of Sect. \ref{sect:scatter_profs_miro_obs_comparison}. The main difference from the previous analysis is that this time we are not able to directly map the positions of the observed RMs in the SLOW SSC analog maps. Therefore, we used the value of the $\RM$ maps inside the entire analog on-target region\footnote{$4\times 10^{-5} \leq n_{e} \, \text{\small [cm$^{-3}$] } \leq 8\times 10^{-4}$, see \ref{sect:slow_method_RM_scatter}.} to produce the profiles. However, we binned the SLOW profiles in the same way as the observational profile to make the comparison with the data by evaluating $\mathfrak{S}_{\RM}^{\text{SLOW}}$ at the same $d_{\text{nrst},i}$ bins (see Appendix \ref{sect:slow_method_scatter_profiles}).

\begin{table}[h]
    \caption{ Bayesian evidence, Bayes factors and K\&R scale result for the SLOW magnetic field amplification mechanisms model selection analysis. }  
    \centering
    \renewcommand{\arraystretch}{1.1}
    \setlength{\tabcolsep}{2.8pt}
    \small{
    \begin{tabular}{c|cccccc}
        Mechanism & $B_{\mathcal{F}}$ & $B_{\text{sim}}$ & $B_{\text{ff}}$ & $B_{\beta}$ & $B_{\text{dyn}, \downarrow}$ & $B_{\text{dyn}, \uparrow}$ \\
        \hline \hline
        $\ln{\mathcal{L}_k}$ & -116 & -121 & -176 & -206 & -266 & -269 \\
        $\ln{\mathcal{B}}$ & ... & 5.3 & 60 & 90 & 151 & 153 \\
        K\&R & ... & V. strong & V. strong & V. strong & V. strong & V. strong
    \end{tabular}
\tablefoot{V. strong stands for very strong evidence against the models with respect to the one with highest Bayesian evidence.}
\label{tab:bayes_slow} }
\end{table}

In Table \ref{tab:bayes_slow} we show the results from the Bayesian analysis. The turbulent scenario for amplification of the magnetic field in the intercluster region, i.e., $B_{\mathcal{F}}$ is the one that provides the best description of the data. We have very strong evidence against all other mechanisms according to the K\&R scale.


\section{Discussion}\label{sect:discussion}

The main objective of this work is to characterise the properties of the detected magnetized gas in the SSC (Sections \ref{sect:sigma_RM_on_target} and \ref{sect:obs_RM_scatter_bridge_clusters}). To do this, we first modeled the SSC with Gaussian random magnetic fields with the MiRò code (Sect. \ref{sect:miro_results}). These models provide insights into the correlation of magnetic field strength and electron density in the ICM through the $\eta$ parameter defined as $\abs{\Vec{B}} \propto n_e^{\eta}$.
However, after estimating $\sigma_{\RM}$ in the bridge and clusters regions predicted by MiRò (Sect. \ref{sect:miro_signals}), we found a model degeneracy, i.e., with this information alone, we are not able to distinguish between models with different values of $\eta$. Moreover, we have also estimated how the uncertainties in $\sigma_{\RM}$ behave when increasing the $\RM$ grid density, and found that even for SKA-expected grid densities $\sim 100$ $\RM$s/deg$^2$, this statistic suffers from large uncertainties (see \ref{sect:RMgrid_dens}).
These statistical difficulties are not model-dependent, i.e., the local-constrained cosmological MHD simulation SLOW (Sect. \ref{sect:slow_results}), despite providing a much more detailed prescription of ICM physics, suffers from them as well (Fig. \ref{fig:slow_obs_signal_comparison}). Therefore, this motivates the need to study the trend in the $\RM$ scatter as a function of distance from the cluster and group centers ($\mathfrak{S}_{\RM}(d_{\text{nrst}})$ profiles) to gain further insight into the physical properties of the magnetized gas in the SSC.

\subsection{Flatter-than-expected $\mathfrak{S}_{\RM}$ profiles. The $\eta<0.5$ puzzle.}

As mentioned in the previous subsection, while Gaussian random field models (MiRò) with $\eta=0.5$, corresponding to a constant thermal-to-magnetic energy scaling, are able to match the observations in terms of $\sigma_{\RM}^{\text{Clusters}}$ (Fig. \ref{fig:miro_obs_signal_comparison}), there are degeneracies with other models with $\eta<0.5$. 

The analysis of the $\RM$ scatter as a function of distance to the nearest cluster/group, $\mathfrak{S}_{\RM}(d_{\text{nrst}})$ (Sect. \ref{sect:scatter_profs_miro_obs_comparison}), revealed that the model with highest Bayesian evidence, hence, the one with best predictive performance, is $B_{\text{rms}}=2.5\,\mu\mathrm{G}$, $\eta=0$, which has an unphysical scaling of $\eta=0$ (Table \ref{tab:bayes_miro}). Furthermore, we found positive evidence against the best-performing $\eta\geq 0.5$ model, i.e., $B_{\text{rms}}=1\,\mu\mathrm{G}$, $\eta=0.5$. In an attempt to overcome the difficulties in the analysis introduced by the complicated geometry and overlapping structures, we analysed profiles with respect to electron density $\mathfrak{S}_{\RM}(n_e)$ and followed the same Bayesian analysis for all MiRò models \ref{sect:RM_scatter_profiles_n_e}. While these further indicate that $B_{\text{rms}}=2.5\,\mu\mathrm{G}$, $\eta=0$ is still the best-performing model, which has the unphysical $\eta=0$ scaling, there is only barely worth mentioning evidence against the $B_{\text{rms}}=1\,\mu\mathrm{G}$, $\eta=0.5$ model (Table \ref{tab:bayes_miro_n_e}).

The observed profile with respect to distance to the nearest cluster/group (Fig. \ref{fig:RM_scatter_obs}), while showing an anti-correlation as suggested by a Spearman rank test, is still considerably flatter than the profiles derived from MiRò, which causes these $\eta<0.5$ models to be consistent with the data. However, the observed profile is consistent with a flat dependence when considering the distance to the nearest cluster only (Sect. \ref{sect:slow_scatter_profiles}), as well as with electron density (Fig. \ref{fig:RM_n_e_scatter_profiles}).  

Considering that MiRò was designed mainly for single-cluster studies, it does not account for interactions between the clusters and groups of the SSC or a spatially variable magnetic field coherence length. 
However, the SLOW simulations overcome these limitations, providing detailed ICM cluster physics with a full interaction and merger history\footnote{Figure \ref{fig:GRM_RRM_maps} shows a potential enhancement of the $\RRM$s to the north of A3562. This could be a tracer of the merger history of the system (e.g., a trail of gas left by SC-1327 if it merged from that direction, as commented on in \citet{venturiRadioFootprintsMinor2022}. A merging event similar to this one is suggested by the SLOW simulated SSC history at $z\sim0.07$.}, amplification of a seed magnetic field by adiabatic compression and dynamo processes throughout the dynamical evolution of the system, and prescriptions for further amplification mechanisms outside the cluster cores. As a consequence, the analog SSC provided by SLOW (Fig. \ref{fig:slow_shapley}) is the best possible description of the system to date. 

The $\mathfrak{S}_{\RM}(d_{\text{nrst}})$ profile comparison between all SLOW amplification mechanisms and the observed profile 
(Fig. \ref{fig:scatter_profiles_SLOW}) indicates, through the Bayesian model selection analysis, that the turbulent velocity scenario ($B_{\mathcal{F}}$) has the best predictive performance (Table \ref{tab:bayes_slow}). There is very strong evidence against all other amplification mechanisms. Nonetheless, there is a large discrepancy at scales $d_{\text{nrst}}\gtrsim 0.8$. It is possible that this could be solved by more directly comparable positions of the analog SC groups with respect to the observed ones (Fig. \ref{fig:slow_shapley} vs Fig. \ref{fig:ShapS_y_RMs}). 

In the $B_{\mathcal{F}}$ scenario, the magnetic field strength is rescaled as a fraction $\mathcal{F}$ of the turbulent pressure, which was assumed to be in equilibrium with magnetic pressure ($\mathcal{F}=1$).
This model defines turbulent velocity as the deviation from the gas bulk velocity \citep[similar to e.g.,][]{Simonte2022} and therefore preferentially amplifies magnetic fields in regions of shocks.
This opens an avenue for future investigations with more elaborate sub-grid models for turbulence \citep[e.g.,][]{Iapichino2017} of magnetic field amplification at shocks via the Bell instability. Recent results by \citet{Zhou2024} show magnetic field generation and amplification in low-density accretion shock regimes which could provide stronger magnetic fields than can be captured in the SLOW simulation.
It will also be worth investigating the impact of astrophysical seeds for magnetic fields such as supernovae and active galactic nuclei (AGN), and how feedback effects contribute to distributing, mixing and amplifying magnetic fields \citep[see e.g.,][for a review]{Donnert2018}. This could have a further impact on the $\eta < 0.5$ puzzle and will need to be investigated with the next generation of SLOW simulations.

Other recent works like \citet{osingaProbingClusterMagnetism2025, khadir25}, have found similar results, where the data suggest models with $\eta < 0.5$. 
Understanding what is driving the flattening of the observed RM profile is key to solving this problem. 
As pointed out by \citet{osingaProbingClusterMagnetism2025}, there seems to be an inherent difference between studies that use resolved radio galaxies, e.g., \citet{murgiaMagneticFieldsFaraday2004, bonafedeComaClusterMagnetic2010, vaccaIntraclusterMagneticField2012, govoniSardiniaRadioTelescope2017, stuardiIntraclusterMagneticField2021}, which find $\eta\geq 0.5$, and those using unresolved radio sources. The latter have found this flatter-than-expected $\mathfrak{S}_{\RM}$ profile as a function of distance from the cluster center \citep{bohringer16, Stasyszyn_delosRios19, osingaDetectionClusterMagnetic2022, osingaProbingClusterMagnetism2025, khadir25}, which favor the $\eta<0.5$ models. The size of the samples is considerably larger in studies where the $\eta<0.5$ models are favored, e.g., \citet{osingaProbingClusterMagnetism2025} stacks a sample of order $\sim100$ clusters with $\sim 600$ $\RM$s, and \citet{khadir25} uses 111 $\RM$. While studies with resolved radio galaxies derive their conclusions from a limited number of sources, typically $\lesssim 10$, their focus on individual clusters allows a more precise control of some of the possible confounding factors in stacking studies and interacting systems \citep[e.g.,][]{bonafedeComaClusterMagnetic2010,pagliottaAbel2142_2025}.  

Note that, the $\RM$ scatter profile derived from observations in this work (Fig. \ref{fig:RM_scatter_obs}), matches the behavior of the profile derived by \citet{osingaProbingClusterMagnetism2025} for sources behind the clusters (see their Fig. 6), i.e., they have the same amplitude of the profile at similar scales and a similar flattening at $\gtrsim 0.7\, r_{500}$. This implies that the behavior found for the SSC profile is not a special case, but rather likely representative of the general population. 

\subsection{On the possibility of a missing RM population due to depolarization}\label{sect:missingRMs}

One major difference between our observations and the SLOW simulations is that the observations suffer from depolarization effects. One expects that sources that probe lower projected distances from the clusters' centers have a large scatter in the $\RM$, and it is at these scales precisely where we could be missing sources due to depolarization. Given our beam's linear size of $\sim 20 \,\mathrm{kpc}$, we cannot resolve spatial fluctuations below this scale. This means that there could be sources with an intrinsic scatter in their $\RM$s that we are not accounting for. We studied this in Section \ref{sect:beam_depol}, to find missing additional scatter of only $\Sigma_{\RM}\lesssim 4$ rad$/$m$^2$, on average, due to this effect. This is clearly not enough to boost the scatter in the first bins of the profiles in Figs. \ref{fig:scatter_profiles} and \ref{fig:scatter_profiles_SLOW}. 

In fact, we can estimate what is the $\Sigma_{\RM}$ scatter we would need to reconcile the observed $\mathfrak{S}_{\RM}(d_{\text{nrst}})$ value in the first bin (much flatter than expected, see Fig. \ref{fig:scatter_profiles_SLOW}) with the value predicted by the $B_{\mathcal{F}}$ magnetic field amplification scenario (the best performing model according to the Bayesian analysis).
We would need 16 sources (the size of the sliding window used to compute the profile) with $\Sigma_{\RM}\simeq 50$ rad m$^{-2}$ to yield a $\mathfrak{S}^{\text{Obs}}_{\RM}(d_1)\simeq \mathfrak{S}_{\RM}^{B_{\mathcal{F}}}(d_1)$, where $d_1$ corresponds to the first bin in $d_{\text{nrst}}$. This implies we would be losing a scatter as high as $\Sigma_{\RM}\simeq 1.65\, \sigma_{\RM}^{\text{SSC}}$. 
Moreover, we do not find any significant difference in the density of the $\RM$ grid in regions closer to the clusters centers than at larger distances. 
Note though that in this work, we set a threshold of $p\ge1\%$ to produce the catalog, while future ASKAP observations of the Shapley Supercluster may allow us to use a lower threshold of $0.1\%$ \citep{gaenslerPolarisationSkySurvey2025}. This could reveal more sources potentially enabling an analysis close to the cluster and group centers.

\section{Conclusions}\label{sect:conclusions}

In this work we have studied the properties of the magnetized gas in the Shapley Supercluster Core (SSC), with a special emphasis on the $\simeq 4.2\,\mathrm{Mpc}$ (projected) intercluster region (or ``bridge'') between the massive Abell clusters (A3558-A3562), by using radio polarization data from the ASKAP-POSSUM Pilot II survey in combination with thermal Sunyaev-Zeldovich effect data (from Planck).

Upon the detection of this magnetized gas, we followed different modeling approaches to characterise this gas, namely: semi-analytic galaxy cluster modeling with universal density profiles and Gaussian random fields following a power spectrum (MiRò), and constrained local-Universe cosmological MHD simulations (SLOW). Both of these models generate mock $\RM$ maps that we used to compare with the $\RM$s we derived from ASKAP-POSSUM observations. The main conclusions derived from this work are the following: 

\begin{enumerate}
    \item We have detected a highly significant ($6.7\sigma$) excess $\RM$ scatter from the magnetized gas present in the SSC, $\sigma^{\text{SSC}}_{\RM} = 30.52\pm 4.55 \,\, \mathrm{rad/m^2}$. Further, we have determined that this excess is not dominated by the Abell clusters' contribution alone, given that the bridge region where the massive groups SC 1327, SC 1329 are, has a comparable excess $\RM$ scatter to the clusters $\sigma^{\text{Bridge}}_{\RM} = 25.32\pm 8.48 \,\, \mathrm{rad/m^2}$. 

    \item We have also studied the trend of the $\RM$ scatter as a function of distance (normalised by $r_{500}$) to the nearest cluster/group $\mathfrak{S}_{\RM}(d_{\text{nrst}})$, in the range sampled by our data $0.3 \lesssim d_{\text{nrst}} \lesssim 1.8$. A Spearman rank test revealed a moderate anti-correlation (correlation coefficient $-0.54$) with a p-value of $2\times 10^{-3}$. Overlapping structures in the SSC appear in our data for $d_{\text{nrst}} > 0.7\,r_{500}$. 
    
    \item We have estimated the average magnetic field strength in the bridge and clusters regions (Fig. \ref{fig:subsamples}). Single-scale model estimates, for a variety of assumptions, yield 0.2-1.7 $\mu$G (Table \ref{tab:B_single_scale}); MiRò estimates are $\simeq1.7\,\mu$G for the best performing model (Sect. \ref{sect:miro_B_avg}), while SLOW predicts 1-3 $\mu$G for all amplification mechanisms, except for $B_{\text{sim}}$, that yields $\sim0.1\,\mu$G.

    \item When comparing MiRò with observations, models with $\eta<0.5$ are preferred by the data, as revealed by comparison of $\RM$ scatter profiles both with distance ($\mathfrak{S}_{\RM}(d_{\text{nrst}})$, Sect. \ref{sect:scatter_profs_miro_obs_comparison}) and electron density ($\mathfrak{S}_{\RM}(n_e)$, Appendix \ref{sect:RM_scatter_profiles_n_e}) through the Bayesian model selection analysis.
    
    \item The SLOW simulations provide the most detailed and complete description of the ICM physics and, despite minor differences in projected distance between the clusters and the position of the groups, an analog SSC system was identified (Fig. \ref{fig:slow_shapley}). Out of the six different amplification mechanisms for the magnetic field, the $\RM$ scatter profile analysis (Sect. \ref{sect:slow_scatter_profiles}) indicates that the best performing model is the one with amplification by turbulent velocity $B_{\mathcal{F}}\propto n_e^{1/2} v_{\text{turb}}$ in the intercluster region, with equilibrium between turbulent and magnetic energies. 

    \item The observational RM profiles are flatter than expected compared to all models considered in this work. We refer to this as the $\eta<0.5$ puzzle, which cannot be easily solved by accounting for depolarization effects (considering the instrumental polarization of the ASKAP Pilot survey observations of $\lesssim1\%$). We do not find evidence for a large  missing population of sources close to the cluster centers that would contribute to the $\RM$ scatter sufficiently (see \ref{sect:missingRMs}).  

\end{enumerate}

Given that the RM profile of the SSC is not a special case, and likely representative of the RM behaviour of the general population (Sect. \ref{sect:discussion}), it provides the perfect laboratory for enhancing our understanding of cluster physics and magnetic field amplification scenarios (e.g., does magnetic reconnection play a significant role?). 
In the near future, the full POSSUM survey data, with an order of magnitude better threshold in the degree of polarization \citep{gaenslerPolarisationSkySurvey2025} will be able to detect more $\RM$s towards the core of the clusters and groups. Deeper MeerKAT, and eventually SKA-Mid, observations can further improve $\RM$ grid densities, while their higher angular resolution  \citep{braunAnticipatedPerformanceSquare2019} should allow for more detailed analysis on scales closer to the cluster centers. Higher resolution tSZ maps from the Simons Observatory \citep{simons_obs} will also significantly improve the electron density mapping of complex systems like the SSC. 

\begin{acknowledgements}

D. Alonso-López would like to thank the GUAIX group at the UCM for useful help and comments throughout the development of this work. D. Alonso-López also acknowledges support from the Universidad Complutense de Madrid and Banco Santander through the predoctoral grant CT25/24 and PID2022-138621NB-I00, funded by MCIN/AEI/10.13039/501100011033/FEDER, EU.
D. Alonso-López and SPO acknowledge support from the Comunidad de Madrid Atracción de Talento program via grant 2022-T1/TIC-23797, and grant PID2023-146372OB-I00 funded by MICIU/AEI/10.13039/501100011033 and by ERDF, EU. 
This scientific work uses data obtained from Inyarrimanha Ilgari Bundara/the Murchison Radio-astronomy Observatory. We acknowledge the Wajarri Yamaji People as the Traditional Owners and native title holders of the Observatory site. The Australian SKA Pathfinder is part of the Australia Telescope National Facility (https://ror.org/05qajvd42) which is managed by CSIRO. Operation of ASKAP is funded by the Australian Government with support from the National Collaborative Research Infrastructure Strategy. ASKAP uses the resources of the Pawsey Supercomputing center. Establishment of ASKAP, the Murchison Radio-astronomy Observatory and the Pawsey Supercomputing Centre are initiatives of the Australian Government, with support from the Government of Western Australia and the Science and Industry Endowment Fund. The POSSUM project (https://possum-survey.org) has been made possible through funding from the Australian Research Council, the Natural Sciences and Engineering Research Council of Canada, the Canada Research Chairs Program, and the Canada Foundation for Innovation.
AB acknowledges support from the ERC-CoG project  BELOVED n. 101169773. KD acknowledges support by the COMPLEX project from the European Research Council (ERC) under the European Union’s Horizon 2020 research and innovation program grant agreement ERC-2019-AdG 882679. 
BAS acknowledges support by the grant agreements ANR-21-CE31-0019/490702358 from the French Agence Nationale de la Recherche/Deutsche Forschungsgemeinschaft, DFG through the LOCALIZATION project. BAS acknowledges support by the COMPLEX project from the European Research Council (ERC) under the European Union’s Horizon 2020 research and innovation program grant agreement ERC-2019-AdG 882679. LMB would like to thank Frederik Groth for help with the vortex-p code. LMB is supported by NASA through grant 80NSSC24K0173. C.S.A. acknowledges funding from the Australian Research Council in the form of FT240100498.
POSSUM is partially funded by the Australian Government through an Australian Research Council Australian Laureate Fellowship (project number FL210100039 awarded to NM-G). The Dunlap Institute is funded through an endowment established by the David Dunlap family and the University of Toronto. Basic research in radio astronomy at the Naval Research Laboratory is supported by 6.1 Base funding. SM acknowledges support from grant PID2023-146372OB-I00, funded by MICIU/AEI/10.13039/501100011033 and by ERDF, EU. 

\end{acknowledgements}

\bibliographystyle{aa}
\bibliography{references}

\begin{appendix}
\section{Observational methods and ancillary results}\label{sect:obs_methods}

\subsection{Stokes $I$ emission of the ASKAP-POSSUM radio galaxies}

\begin{figure*}[ht]
    \centering
    \includegraphics[width=0.8\linewidth]{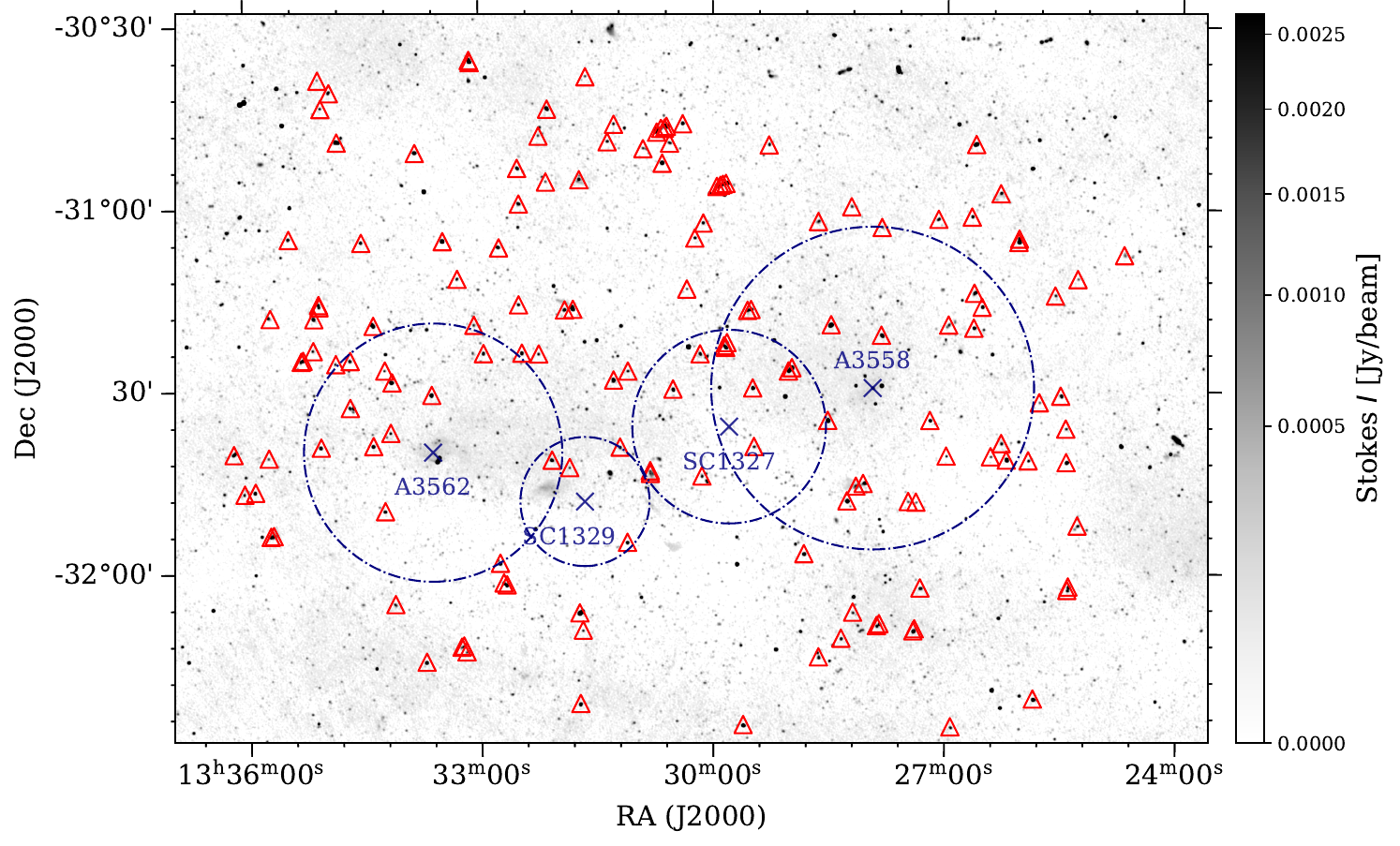}
    \caption{Stokes $I$ image of the ASKAP fields with the location of the 149 final $\RM$s in the catalog and the centers of the clusters and groups of the Shapley Supercluster core relevant for this work. The circles represent their $r_{500}=1.5,\,1.2, \, 0.9,\,0.6\,\, \mathrm{Mpc}$. The colorbar is in units of $\mathrm{Jy/beam}$. }
    \label{fig:StokesI_RMs}
\end{figure*}

Figure \ref{fig:StokesI_RMs} shows the Band 1 ($800$-$1088$ MHz) Stokes $I$ image from the ASKAP-POSSUM observations used to derive the $\RM$ grid of this work. 

\subsection{Robustness of GRM estimation and structure function analysis}\label{sect:grm_tests}

We have performed a series of tests on mock $\RM$ data for the full 2475 sources used during the $\GRM$ estimation to determine the robustness of the annulus method. The mock $\RM$s were divided into the on-target region, and the rest of the ``core'' and ``south'' POSSUM fields (which we will denote by not-on). The mock $\RM$s of this not-on distribution, were drawn from a Gaussian distribution with constant standard deviation $\sigma^{\text{not-on}}_{\RM}=12$ rad/m$^2$ over the whole field  The mock on-target distribution, was further split into two bins based on their $d_{\text{nrst}}$ values. Those sources with $d_{\text{nrst}}\leq$ than the median distance, were drawn from a Gaussian distribution with $\sigma^{\text{on, below}}_{\RM}=40$ rad/m$^2$ while those at $d_{\text{nrst}}>$ than the median from another Gaussian with $\sigma^{\text{on, above}}_{\RM}=20$ rad/m$^2$. 
The mean is common for all three input distributions $\langle\RM\rangle=20$ rad/m$^2$.

To test the performance of the annulus method, we analysed to what level the standard deviations of the input mock $\RM$s are recovered by the final $\RRM$s. We parametrize this with a deviation parameter that reflects the change in the standard deviation before and after applying the annulus method:
\begin{equation}
    \epsilon\equiv1-\frac{\sigma_{\RM}}{\sigma_{\RRM}}.
\end{equation}
Thus, $\abs{\epsilon}\simeq 0$ would indicate a good performance of the annulus method. However, we need to take into account the statistical errors computed as the standard error of the standard deviation ($\delta\sigma_{\RRM}$). The relative error $\Delta\sigma\equiv\delta\sigma_{\RRM}/\sigma_{\RRM}$ is then completely determined by the number $N$ of mock $\RM$s of each sub-sample $\Delta\sigma=(2(N-1))^{-1/2}$. These errors go from $7\%$ to $15\%$ for $N=103$ (off-target) and $N=23$ (on-target split in two halves), respectively. 
Therefore, to test, within the statistical errors the performance of the annulus method on recovering the input statistics we need to evaluate the quotient $\abs{\epsilon}/\Delta\sigma$. 

We ran the simulation $n=1000$ times, and obtained a median deviation parameter of $\abs{\epsilon} = 0.05$, thus indicating a great performance of the annulus method. The median quotient $\abs{\epsilon}/\Delta\sigma$ goes from 1.4-3.2 for the different sub-samples, which implies that the relative errors are typically a factor $\lesssim 3$ lower than the deviation parameter. Therefore, the results of this robustness test are not dominated by the statistical uncertainties. 

Then, we tested by what percentage the annulus GRM method can flatten the $\mathfrak{S}_{\RM}(d_{\text{nrst}})$ profile by checking the relative slope between the sources in the two $d_{\text{nrst}}$ bins before (input mock $\RM$s), and after applying the annulus ($\RRM$s). After running the simulation $n=1000$ times, we found the annulus $\GRM$ method to flatten the $\RM$ scatter profile by $\sim 0.3\%$. This result illustrates the robustness of our $\GRM$ modeling as well as proving that it does not play any significant role in flattening the observed $\RM$ profiles of Sect. \ref{sect:obs_RM_scatter_prof}.

We also carried out a structure function analysis\footnote{\href{https://github.com/AlecThomson/structurefunction}{https://github.com/AlecThomson/structurefunction}} \citep{thomsonStructureFunction2024} on the residual rotation measures. It revealed a flat behavior with respect to the angular size on scales between $0.2$-$2$ degrees. This means that for pairs of sources whose projected separation is different, no significant variation is found in the difference of their $\RM$s. This is interpreted as a successful foreground removal. A Galactic-type contribution in the data, is associated with a non-flat structure function, since large gradients of coherent patches expected from it would make the structure function peak at angular scales across them \citep{haverkornMagneticFieldsIonized2004}.

\subsection{Path length estimation}\label{sect:L}
    One of the key parameters in the single-scale model approach is the path length $L$. This represents the extent of the Shapley Supercluster that Faraday rotates the angle of polarization of the light coming from our background sources. Thus, to properly determine its value, one needs to assume certain aspects of Shapley's geometry. We have followed a simple approach, where we have assumed that both A3558 and A3562 as well as the two galaxy groups between them are at the same redshift, i.e., they lie on the same plane. We have further assumed that the bridge is axisymmetric, and the axis of symmetry is parallel to the plane of the sky. Then, we can assume that the depth is equal to the width of the projection of the bridge in the plane of the sky.

\begin{figure}[ht!]
    \centering
    \includegraphics[width=1\linewidth,clip=true,trim=0cm 1cm 0cm 3cm]{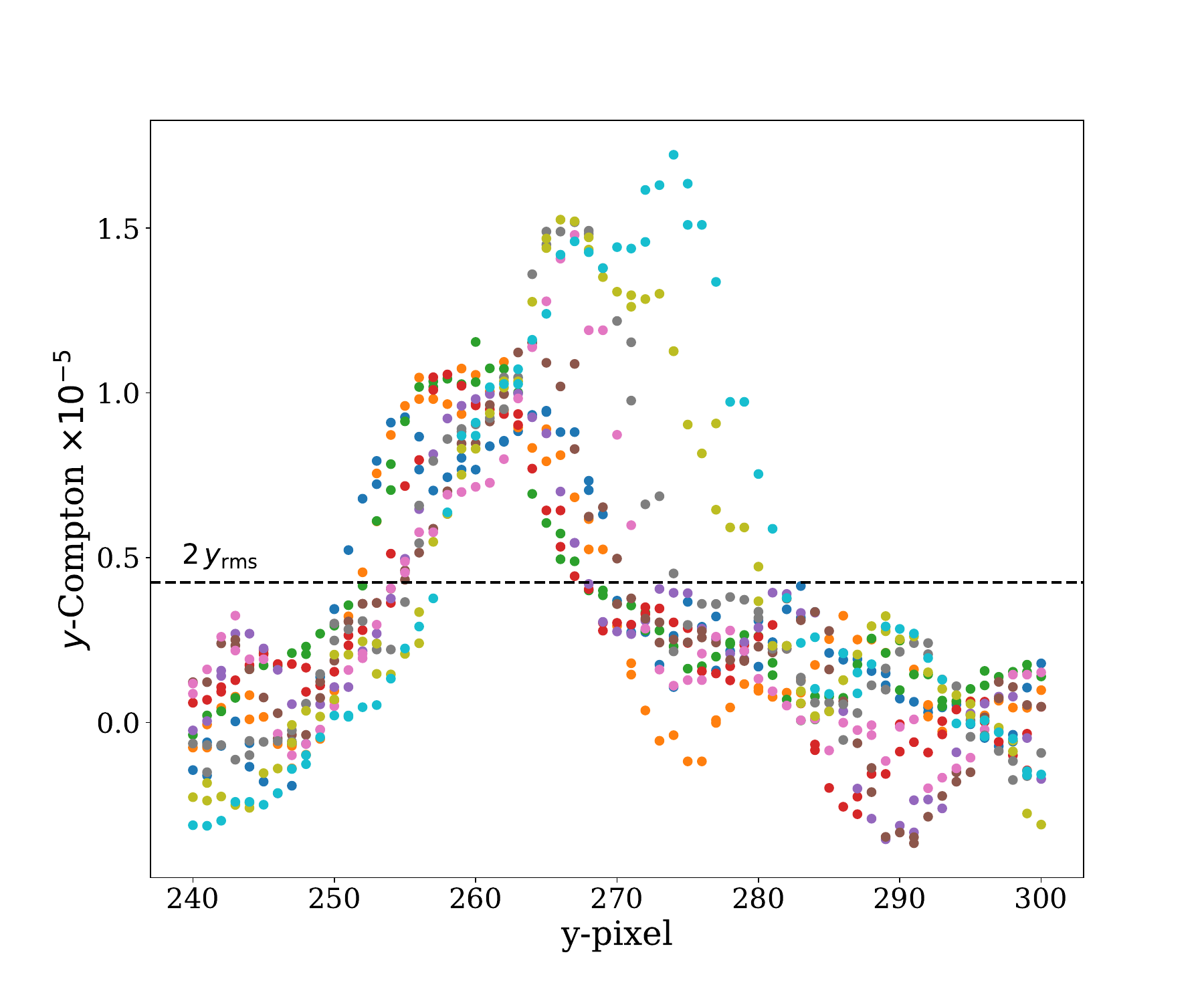}
    \caption{Transverse $y$-map profiles for the bridge. 1 pix$=1.17$'. The horizontal line represents $y_{\text{bdry}}$. For each profile, the difference between where the two values that cut this line represent the estimation for the path length. The resulting value of $L\simeq1.8\,\mathrm{Mpc}$ is the median of all transverse profiles.}
    \label{fig:path_length_bridge_ymap}
    \end{figure}

    Under these assumptions, we have estimated $L$ by studying transverse profiles through the bridge in the $y$-map. In particular, the axis of symmetry is parallel to the line joining the centers of A3558 and A3562, so we have examined profiles in lines perpendicular to that. Then, the value for $y_{\text{bdry}}$ defines the width of the bridge (Figure \ref{fig:path_length_bridge_ymap}). The median of all the estimated path lengths yields the value $L=(1.8\pm 0.1)\,\mathrm{Mpc}$ used throughout this paper.
    
    This approach gives a result for the path length of lines of sight that go through the bridge. However, we have also used the value derived in this way to estimate the magnetic field strength using the single-scale model (Section \ref{sect:single_scale}) for the clusters sub-sample. 
    Note that, using Eqn. \eqref{eq:n_e_obs} we can see that the ratio between electron densities estimated with different path lengths, say, $L$ and $\tilde{L}$, is $n_e(L)/n_e(\tilde{L}) = \tilde{L}/L$. Then, from Eqn. \eqref{eq:B_single_scale} it follows that the magnetic field strength ratio is
    \begin{equation}
        \frac{B(L)}{B(\tilde{L})} = \sqrt{\frac{L}{\tilde{L}}}\equiv \mathcal{R}_B(L,\tilde{L}).
        \label{eq:Brms_ratio}
    \end{equation}
    Then, once we estimate what is the path length through the clusters $\tilde{L}$, we can determine the robustness of our assumption by evaluating Eqn. \eqref{eq:Brms_ratio}. Following the same method as the one outlined before for the bridge, we have found that the (median) path lengths for the clusters are: $\tilde{L}_{\text{A}3558}=2.6\,\mathrm{Mpc}$ and $\tilde{L}_{\text{A}3562}=2.3\,\mathrm{Mpc}$, which gives ratios $\mathcal{R}_B(1.8,2.6)=0.86$ and $\mathcal{R}_B(1.8,2.3)=0.88$, respectively. It should be clear then, that for a given polarized source in the clusters sub-sample, we expect a factor $\approx 0.9\sim1$ difference in the rms magnetic field strength estimate using a single-scale model, which is not a significant effect and will not affect any conclusions derived from our results.

    We have explored yet another way of path length estimation via the universal density profile cube we created (see \ref{sect:miro_RM_scatter_method}). In this case, we have made use of the convenience of having a 3D cube and we have directly extracted the density profile along the lines of sight of our sources: $n_e(x_s,y_s,z)$, where $x_s,y_s$ are the plane-of-the-sky coordinates of a given source in the detected sample. 
    
    \begin{figure}[ht!]
    \centering
    \includegraphics[width=1\linewidth,clip=true,trim=0cm 1cm 0cm 3cm]{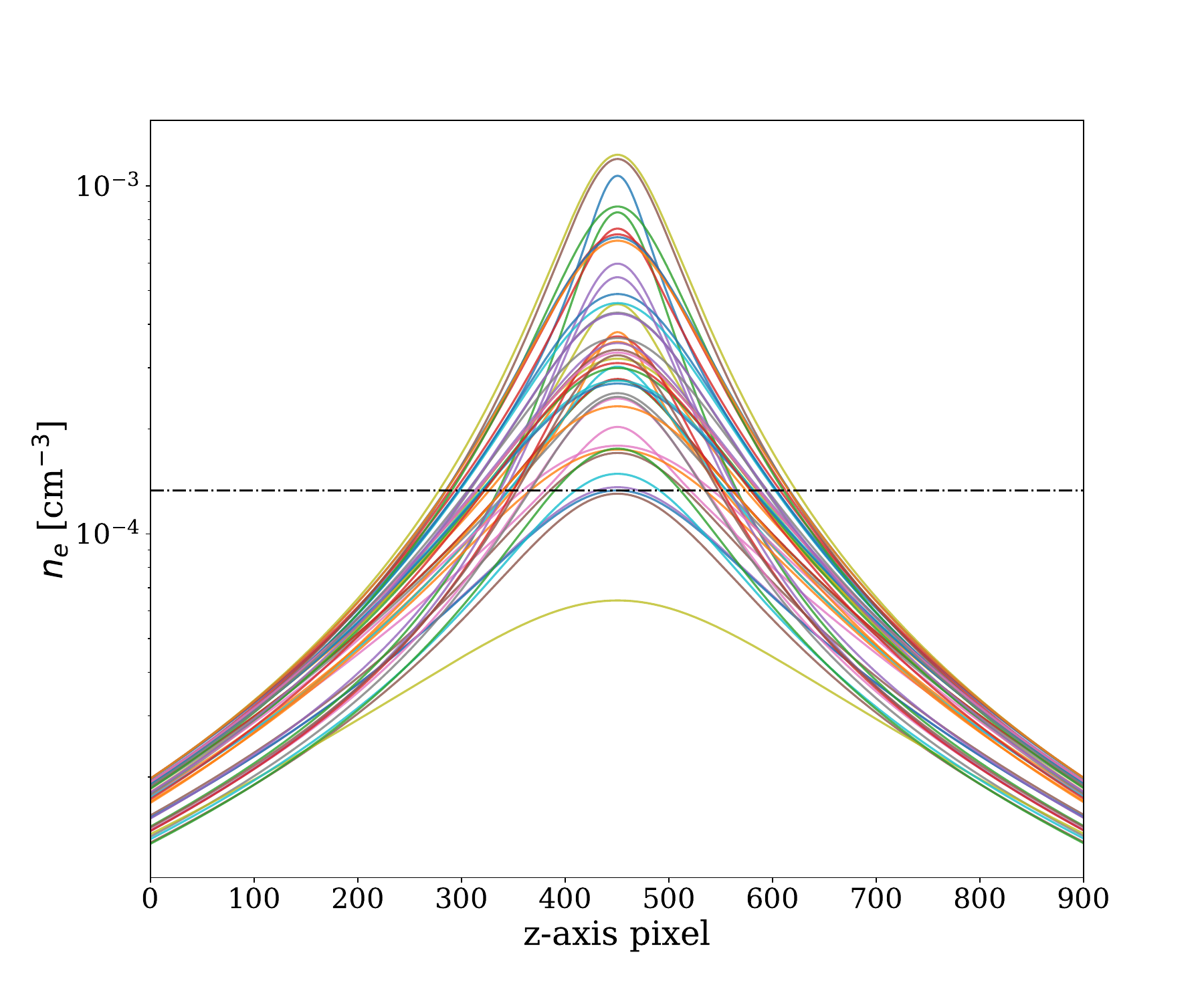}
    \caption{Line of sight universal density profiles through our sources in the on-target sample. The horizontal line represents $n_{e,{\text{thresh}}}=1.33\times 10^{-4}\,\mathrm{cm}^{-3}$. The mean path length for the clusters is $L_{\text{clust}} \simeq 2.6\,\mathrm{Mpc}$, while for the bridge $L_{\text{bridge}} \simeq 2.1\,\mathrm{Mpc}$, just like with the y-map transverse profiles.}
    \label{fig:path_length_bridge_n_e}
    \end{figure}

    To estimate the path length from these, we proceeded in the same way as before, by setting a threshold\footnote{We used $n_e=1.33\times 10^{-4}\,\mathrm{cm}^{-3}$, which is the density value associated with $y_{\text{bdry}}$ assuming a constant temperature of $T=5\times 10^{7}\,\mathrm{K}$.}, and estimating the depth of each line of sight as the width of the profile above the threshold.
    We found that the median bridge path length is $L_{\text{bridge}}=(2.1\pm 0.3)\,\mathrm{Mpc}$, while for the clusters $L_{\text{clust}}=(2.6\pm 0.2)\,\mathrm{Mpc}$. These values are similar to what we found with the $y$-map method. These results indicate that our path length estimations are consistent with each other.

\subsection{X-ray counts and temperature maps from XMM-Newton}\label{sect:xmm}

We have used a temperature map and an X-ray counts map, both derived from XMM-Newton observations (proposal IDs 060198 $\&$ 065159, PI Herve Bourdin). The counts map has a pixel size of $0.11'$. The temperature map is the same as in \citet{venturiRadioFootprintsMinor2022}, made with a multi-scale spectral mapping algorithm by means of a B2-spline wavelet analysis \citep{Bourdin2008}, but smoothed using the tophat function with radius of 20 pixels (132'') in DS9. The pixel size of this map is then $0.23'$. The smoothing was done to fill the holes in the map that come from removing emission from AGNs that do not belong to Shapley. The particular scale of $132''$ corresponds to $\sim 135\,\mathrm{kpc}$, that ensures we minimise small scale fluctuations in the map. This temperature map was used in Appendix \ref{sect:obs_n_e_degeneracy} to estimate the electron density of the SSC.

\subsection{Breaking the $n_eB_{\parallel}$ degeneracy using tSZ and X-ray}\label{sect:obs_n_e_degeneracy}

From the $\RM$ information alone, we can only probe the product of density of free thermal electrons with the magnetic field strength parallel to the line of sight (Eqn. \eqref{eq:RMobs}); thus, further information is needed to estimate the density in Shapley before handling the magnetic field strength. From Eqn. \eqref{eq:ycompton}, we see that we can determine $n_e$ after assuming a dependence with line of sight distance of density and temperature. The most simple approach is to assume that density and temperature are constant throughout a given path length $L$ across Shapley. We estimated the path length using transverse profiles of the $y$-map through the bridge and obtained $L=1.8\,\mathrm{Mpc}$ (see Appendix \ref{sect:L}).
Then, for a given polarized source we can estimate its electron density simply as
\begin{equation}
    n_e=\frac{m_ec^2}{\sigma_Tk_B}\frac{y}{T_eL}.
\label{eq:n_e_obs}
\end{equation}
In particular, using Eqn. \eqref{eq:n_e_obs} and the path length $L$, we have employed four different approaches to estimate the density, three of them based on the XMM temperature map (Sect. \ref{sect:xmm}), and one of them using the universal density cube we created (see \ref{sect:univ_profiles}):
\begin{enumerate}[label=\alph*.]
    \item \underline{$T=10^{7}$ K}: we assume a constant temperature throughout the path length $L$ of $10^{7}$ K. It corresponds to the average of the temperature map. This is a lower limit of the average temperature in the ICM, thus giving an upper limit for the electron densities. 
    \item \underline{$T=5\times 10^{7}$ K}: same method as above but with a temperature five times higher, which is the average of the temperature map at the location of the on-target sources. 
    \item \underline{XMM $T$-map}: we use the exact value of the temperature map at the location of each of the on-target sources.  
    \item \underline{Universal $n_e$ profiles}: we evaluate the projected density map generated from the universal density profiles for each object (see Section \ref{sect:univ_profiles}) at the position of the on-target sources.
\end{enumerate}

\begin{table}[ht!]
\caption{ Range of thermal electron density values probed by each sub-sample as well as their median value $\Bar{n}_e$. }
\centering
\renewcommand{\arraystretch}{1.1}
\begin{tabular}{c|ccc}
Sub-sample & Sources & Min-Max $n_e$ [cm$^{-3}$]& $\Bar{n}_e$ [cm$^{-3}$] \vspace{1pt} \\ \hline \hline
   On-target & 46 & 1.73-9.45 $\times 10^{-4}$ & 3.38$\times 10^{-4}$ \\
   Bridge & 12 & 1.77-5.24 $\times 10^{-4}$ & 3.36$\times 10^{-4}$ \\
   Clusters & 34 & 1.73-9.45 $\times 10^{-4}$ & 3.78$\times 10^{-4}$
\end{tabular}
\tablefoot{All densities are computed with the individual temperature values of the XMM map at the position of the on-target sources using Eqn. \eqref{eq:n_e_obs} (method (c)).}
\label{tab:obs_n_e}
\end{table}

All of these methods were used to estimate the magnetic field strength from a single-scale model (see Sect. \ref{sect:single_scale}, Appendix \ref{sect:single_scale_details} and Table \ref{tab:B_single_scale}). To define the on-target region in MiRò we related $y_{\text{bdry}}$ with an electron density value to define the $\mathscr{N}_e$ regions using method (c) (see Appendix \ref{sect:miro_RM_scatter_method}). The resulting $n_e$ range that we can probe in all three observed sub-samples as obtained with method (c) can be found in Table \ref{tab:obs_n_e}.

\subsection{Single-scale magnetic field model details}\label{sect:single_scale_details}

\begin{table*}[ht!]
\small{
\caption{Magnetic field strength estimates from the observations assuming a single scale model (Sect. \ref{sect:single_scale} and \ref{sect:single_scale_details}). }
\label{tab:B_single_scale}
\centering
\renewcommand{\arraystretch}{1.25}
\begin{tabular}{c|c|c|c|c|c|c}

 & \multicolumn{3}{c|}{$\Bar{n}_e$ [cm$^{-3}$]} & \multicolumn{3}{c}{$B$ [$\mu$G]} \\ \cline{1-7} 

Method & SSC & Bridge & Clusters & SSC & Bridge & Clusters \\ \hline \hline

$T=10^{7}$ K & $1.27\times 10^{-3}$ & $1.55\times 10^{-3}$ & $1.16\times 10^{-3}$ & $0.34\pm0.11$ & $0.23\pm 0.09$ & $0.33\pm 0.12$ \\
$T=5\times 10^{7}$ K & $2.54\times 10^{-4}$ & $3.10\times 10^{-4}$ & $2.31\times 10^{-4}$ & $1.71\pm 0.53$ & $1.16\pm 0.47$ & $1.67\pm 0.58$ \\ 
XMM $T$-map & $3.38\times 10^{-4}$ & $3.36\times 10^{-4}$ & $3.78\times 10^{-4}$ & $1.28\pm 0.39$ & $1.07\pm 0.43$ & $1.02\pm 0.35$ \\ 
Universal $n_e$ & $2.62\times 10^{-4}$ & $2.26\times 10^{-4}$ & $2.78\times 10^{-4}$ & $1.65\pm 0.51$ & $1.60\pm 0.64$ & $1.39\pm 0.48$ \\ 
\end{tabular}
\tablefoot{The $\Bar{n}_e$ values correspond to the median of all observed sightlines through a given sub-sample as estimated by the methods in Section \ref{sect:obs_n_e_degeneracy}. The coherence length is fixed to $\Lambda=50\,\mathrm{kpc}$. The assumed path length for all sub-samples is $L=1.8\,\mathrm{Mpc}$ (see Appendix \ref{sect:L}). }
}
\end{table*}

One of the most used models to estimate the magnetic field strength in clusters assumes an idealised scenario where the ICM is split into cells of equal size, with a constant electron density $\Bar{n}_e$ throughout the path length $L$. In each of these cells, of size $\Lambda$, the magnetic field and the line of sight vector form a random angle ($\cos{\theta}$) drawn from a uniform distribution between $-1$ and $1$ (see e.g., \citet{gaenslerRadioPolarizationInner2001a} and references therein).
For a given line of sight, provided that $\Lambda \ll L$, i.e., for sufficiently many number of cells N, summing over the random angles yields a expected value for the rotation measure of $\langle\RM\rangle=0$. Therefore, the $\RM$ grid will be Gaussian distributed as a consequence of the Central Limit Theorem. The standard deviation of this distribution is found to be Eqn. \eqref{eq:sigmaRM}. 
The factor $(2\sqrt{3})^{-1}$ has two origins: $1/\sqrt{3}$ comes from the Gaussianity of the magnetic field distribution, so that, on average,  $B_i=\abs{\Vec{B}}/\sqrt{3}$, while $1/2$ is introduced in \citet{gaenslerRadioPolarizationInner2001a} as a result of considering the possibility of lines of sight with randomly aligned cells. 

We have fixed the coherence scale to $\Lambda = 50\,\mathrm{kpc}$ \citep{guidettiIntraclusterMagneticField2008, bonafedeComaClusterMagnetic2010, vaccaIntraclusterMagneticField2012, govoniSardiniaRadioTelescope2017}, that satisfies the $\Lambda \ll L$ assumption. Finally, we used the electron density estimates in Sect. \ref{sect:obs_n_e_degeneracy} to use Eqn. \eqref{eq:B_single_scale} to compute the magnetic field strength. In particular, using all four methods in that Section, we computed the thermal electron density for each observed $\RM$, and used the median value of each sub-sample as our estimate of $\Bar{n}_e$.

In Table \ref{tab:B_single_scale} we present the magnetic field estimate results from all of the mentioned methods and assumptions. The lower and upper bounds are:
    $0.34\,\mu\mathrm{G}\lesssim B_{\text{SSC}}\lesssim 1.71\,\mu\mathrm{G}$, 
    $0.23\,\mu\mathrm{G}\lesssim B_{\text{Bridge}}\lesssim 1.60\,\mu\mathrm{G}$,
    $0.33\,\mu\mathrm{G}\lesssim B_{\text{Clusters}}\lesssim 1.67\,\mu\mathrm{G}$. Methods (a) and (b) for the $n_e$ estimation provide the lower and upper bound for the SSC and clusters sub-samples, respectively. While in the bridge, method (a) provides the lower bound and method (d) provides the upper bound.
    
The assumed path length $L=1.8\,\mathrm{Mpc}$ was estimated by studying transverse profiles through the bridge in the $y$-map (Appendix \ref{sect:L}). It is worth exploring how variations in the path length can affect our estimates of the magnetic field. For instance, we can think of a source in the clusters sub-sample and the possibility that the path length through the clusters may be larger than the one through the bridge. Then, the magnetic field estimate would be affected by a factor $B(L)\approx 0.9\,B(\tilde{L})$ (Appendix \ref{sect:L}), which is order $1$, where $L$ is the bridge path length and $\tilde{L}\simeq 2.45\,\mathrm{Mpc}$ is the average path length of the clusters. Thus, we have used the estimated path length for the bridge in all cases.

\subsection{Polarized sources catalog}

\begin{table*}[ht!]
\caption{Example of the some of the most relevant data columns to be found in the catalog, for a sample of three sources.  }
\centering
\begin{tabular}{c|c|c|c|c|c|c}
      \multicolumn{1}{c|}{RA [hms]} &
      \multicolumn{1}{c|}{Dec [dms]} &
      \multicolumn{1}{c|}{$\RM$ [rad/m$^2$]} &
      \multicolumn{1}{c|}{$P$ [mJy/\text{beam}]} &
      \multicolumn{1}{c|}{$P/I$ [$\%$]} &
      \multicolumn{1}{c}{Band} &
      \multicolumn{1}{c}{$\RRM$ [rad/m$^2$]} \\
    \hline\hline
      13:30:37.2 & -30:46:58 & -24.16 $\pm$ 0.44 & 1.48 $\pm$ 0.02 & 14.25$\pm$ 0.25 & 1 & -8.29$\pm$3.13 \\
      13:28:39.0 & -31:02:21 & -28.59 $\pm$ 3.75 & 0.13 $\pm$ 0.02 & 2.74 $\pm$ 0.43 & 1 & -8.05$\pm$5.30 \\
      13:30:33.3 & -30:49:28 & -24.47 $\pm$ 1.19 & 0.44 $\pm$ 0.02 & 9.09 $\pm$ 0.44 & 1 & -7.66$\pm$3.48 \\
      ...        & ...       & ...               & ...             & ...             & ... & ... \\
    \end{tabular} 
    \tablefoot{$P\equiv\sqrt{Q^2+U^2}$ is the polarized intensity; $P/I$ is the degree of polarization.}
    \label{tab:master_table}
    \end{table*}
In Table \ref{tab:master_table} we present a sample of three polarized sources of our final $\RM$ catalog with some of the most relevant columns.


\section{Modeling methods and ancillary results}\label{sect:model_methods}

\subsection{Modeling Faraday Rotation from clusters with MiRò}\label{sect:miro_methods}

In galaxy clusters, turbulence is mainly driven by gas flows caused during mergers. Given that the ICM is ionised, the magnetic field gets twisted by the turbulent gas motions (small-scale dynamo), hence, its orientation has an inherent random nature \citep{Donnert2018}. It is natural then to model the magnetic field as a random field. A single line of sight will encounter then many reversals of the field's orientation. For a sufficiently large path length $L$ through the cluster, with enough reversals on scales much smaller $\ll L$, it is equally likely to encounter magnetic fields pointing towards us and away from us, averaging out to zero. Therefore, because of the Central Limit Theorem, the distribution of $\RM$s tends to a Gaussian with zero mean, thus Gaussian random fields are the most common way to model the magnetic field distribution in galaxy clusters e.g., \citep{murgiaMagneticFieldsFaraday2004, bonafedeComaClusterMagnetic2010, bonafedeMeasurementsSimulationFaraday2013, stuardiIntraclusterMagneticField2021, osingaDetectionClusterMagnetic2022, osingaProbingClusterMagnetism2025, derubeisMagneticFieldsOutskirts2024}. A widely used way to generate mock $\RM$ maps that can be compared with observations is with the MiRò code \citep{bonafedeMeasurementsSimulationFaraday2013}. This code implements these Gaussian magnetic fields on top of a three dimensional density cube.

\subsubsection{Universal density profiles}\label{sect:univ_profiles}

As previously mentioned, it is important to map the electron density distribution for a robust estimation of the magnetic field. Therefore, we created a $(9\,\mathrm{Mpc})^3$ density cube, the size of which was chosen to be comparable to that covered by the on and off-target regions colectively, which we denote by $\mathcal{A}$ (see Fig. \ref{fig:subsamples}). The pixel size is $10\,\mathrm{kpc}$. We did a 1:1 match of the relative geometry (at the plane of the sky\footnote{We have assumed the four objects to be at the same redshift which means their centers lay in the $xy$-plane.}) of A3558, A3562 and the two massive groups SC 1329, SC 1327, and assigned each of these objects a universal density profile following \citet{prattLinkingUniversalGas2022}. The main advantage of these profiles is their universality, which accounts for departures from self-similarity by introducing a mass and redshift dependence. These profiles are scaled with the $r_{500}$ of each cluster/group, and since we work at a fixed redshift of $z=0.048$, the only physical parameter related to the clusters that remains free is their mass (see Sect. \ref{sect:intro} to find their values). Defining $x\equiv r/r_{500}$, then the universal density profile follows the equation \citep{prattLinkingUniversalGas2022}
\begin{equation}
    n_e(x,z,M_{500}) = \frac{\rho_{\text{gas}}(x,z,M_{500})}{\mu_em_p} = \frac{\rho_{500}(z)}{\mu_em_p}A(z,M_{500})f(x),
\end{equation}
where $\mu_e=1.148$ is the mean molecular weight per free electron, $m_p$ is the proton's mass and $\rho_{500}(z)=500\rho_c(z)\propto E(z)^2=\Omega_m(1+z)^3+\Omega_{\Lambda}$. The departure from self-similarity function $A(z,M_{500})$ encodes how the self-similarity of these profiles breaks down at larger radii \citet{prattLinkingUniversalGas2022}. The function $f(x)$ follows a generalized Navarro-Frenk-White (GNFW) shape
\begin{equation}
    f(x) = \frac{f_0}{(x/x_s)^{\alpha}(1+(x/x_s)^{\gamma})^{(3\beta-\alpha)/\gamma}}, 
\end{equation}
and the scaling parameters where fixed to the results obtained in \citep{prattLinkingUniversalGas2022}. Note that the cube we generated implicitly assumes non-interacting clusters and groups, i.e., $n_{e,\text{tot}}=\sum_{i=1}^4 n_{e,i}$. 

\subsubsection{RM maps with MiRò: magnetic fields}\label{sect:miro_B_methods}

To model the magnetic field distribution we used Gaussian random fields as implemented by the MiRò code. A brief summary of how the code works is:
\begin{itemize}
    \item To model a Gaussian random magnetic field we work in Fourier space, where the relation between Fourier modes ($k$, in pixels) and spatial scales ($\Lambda$) is given by $k=\ell_{\text{cube}}/\Lambda$, where $\ell_{\text{cube}}$ is the full 1D size of the cube computed as the product of the number of pixels times the resolution.
    \item To make sure that the final magnetic field is divergenceless, we start from the vector potential $\Vec{A}(k)$, whose modulus is drawn from a Rayleigh distribution, while the phase is drawn from a uniform distribution between $0$ and $2\pi$. Then, the magnetic field is $\Vec{B}(k)=i\Vec{k}\times\Vec{A}(k)$. The code makes sure that the output magnetic field follows a given (input) power-spectrum.
    \item Afterwards, everything is transformed back to real space.
\end{itemize}

The total number of pixels of our simulation is $900$ and the pixel size is $10\,\mathrm{kpc}$, chosen such that Nyquist theorem ensures that the minimum fluctuating scale will be $20\,\mathrm{kpc}$, to match the resolution of our ASKAP-POSSUM observations. 

The power spectrum we used was from \citet{dominguez-fernandezDynamicalEvolutionMagnetic2019}, which is motivated by MHD simulations of clusters. The analytical expression for the magnetic energy spectrum is\footnote{This power spectrum is defined between $k_{\text{max}}\propto (20\,\mathrm{kpc})^{-1}$ and $k_{\text{min}}\propto (1\,\mathrm{Mpc})^{-1}$.}
\begin{equation}
    E(k)=A\, k^{3/2}\left[1-\text{erf}\left(B\,\ln{\left(\frac{k}{C}\right)}\right)\right],
    \label{eq:dom-fer}
\end{equation}
where $A$ is a normalization parameter, $B$ sets the width of the spectrum and $C$ is a characteristic wavenumber that corresponds to the inverse outer scale. We have fixed these parameters to $B=1.054$, $C=4.354\,\mathrm{Mpc}^{-1}$, which correspond to the case of a merging cluster at $z=0$, whose power spectrum peaks at $\sim 230\,\mathrm{kpc}$\footnote{To implement this in MiRò, one has to input the $C$ parameter in pixels as $C [\text{pix}]=\ell_{\text{cube}}/230\,\mathrm{[kpc]}=9000\,\mathrm{[kpc}]/230\,\mathrm{[kpc]}$.} (see \citet{dominguez-fernandezDynamicalEvolutionMagnetic2019}). There are two free parameters in our modeling:
\begin{itemize}
    \item $B_{\text{rms}}$: this parameter sets the overall normalization of the magnetic field cube (fixing the value of $A$ in Eqn. \eqref{eq:dom-fer}). Since the magnetic field is Gaussian, we can completely characterize it by its mean, and its standard deviation. For each spatial component, the mean is set to zero $\langle B_i\rangle=0$, that follows from the process outlined before by which the field was produced, and the standard deviation is related to the root-mean-square, which is fixed as an input in the code satisfying: $\sigma^2_{\Vec{B}}=\sum_{i=1}^3\langle B_i^2\rangle= 3\,\sigma_{B_i}^2=B^2_{\text{rms}}$.
    \item $\eta$: it is widely assumed in $\RM$ studies of clusters that the magnetic field scales somehow with density, i.e., $\abs{\Vec{B}}\propto n_e^{\eta}$. Some typical values for this parameter are $\eta=0.5$, corresponding to a thermal-to-magnetic constant ratio scenario (predicted by dynamo models); the conservation of magnetic flux (flux-freezing, see e.g., \citet{klein_fletcher}) and adiabatic compression, lead to $\eta=2/3$ in the case of 3D compression, or $\eta=1$ in the case of 1D compression (e.g., merger shocks), which leads to higher magnetic field strengths in the central regions of the cluster with a steeper radial decline. See Table 6 in \citet{stuardiIntraclusterMagneticField2021} for a summary of the $\eta$ values found in the literature in $\RM$ studies of clusters. 
\end{itemize}

With these inputs and these two free parameters, we have explored the following parameter space: $B_{\text{rms}}= 1\,\mu\mathrm{G}-3.5\,\mu\mathrm{G}$ (in steps of $0.5\,\mu\mathrm{G}$), $\eta=0,\,0.5,\,1$. This particular choice was made to test the aforementioned scenarios of magnetic field correlation with electron density and to find models that reproduce the observed scatter in the clusters region (Sect. \ref{sect:obs_RM_scatter_bridge_clusters}). 

The code outputs three magnetic field cubes $(B_x,B_y,B_z)$ for each realization and a mock $\RM$ map computed by integrating the density and magnetic field cubes along the $z$ axis. 

\subsubsection{Measuring the excess $\RM$ scatter with MiRò}\label{sect:miro_RM_scatter_method}

From the observations we can only sample the magnetic field distribution from an $\RM$ grid, i.e., a statistical sample of background radio sources with individual $\RM$s that probe a limited number of lines of sight through the extended Faraday screen of the Shapley supercluster. 

We want to compare the observed $\sigma_{\RM}^{\Omega}$ found in Sect \ref{sect:obs_RM_scatter_bridge_clusters} with the predictions from MiRò. Here, $\Omega\subset\mathcal{A}$ refers to a 2D region of the maps, either the bridge or clusters region. First, we need to define these regions in the MiRò maps. This is carried out by means of two different criteria: matching the $n_e$ range constrained by observations (Sect. \ref{sect:obs_n_e_degeneracy}, Tab. \ref{tab:obs_n_e}), and defining the same bridge box as in Figure \ref{fig:ShapS_y_RMs}.

\begin{itemize}
     \item To map the observed electron density range into the MiRò analysis, we first projected the density cube by integrating along the $z$-axis over the path length we determined observationally: $L=1.8\,\mathrm{Mpc}$ (Sect. \ref{sect:L}): 
    \begin{equation}
        n_{e,\text{proj}}(x,y) = \frac{1}{L}\sum_{z=z_0-L/2}^{z=z_0+L/2} n_e(x,y,z),
    \label{eq:proj_n_e}
    \end{equation}
    where $z_0$ is the $z$-axis pixel where the clusters and groups centers are set to in the cube. Then, the projected density boundaries are defined in the following way:
        \begin{itemize}
           \item $\mathscr{N}_{e,\text{Bridge}}\equiv 1.33\times 10^{-4} \leq n_{e,\text{proj}} \, \text{\small [cm$^{-3}$] } \leq 5.24\times 10^{-4}$
           \item $\mathscr{N}_{e,\text{Clusters}}\equiv 1.33\times 10^{-4} \leq n_{e,\text{proj}} \, \text{\small [cm$^{-3}$] } \leq 9.24\times 10^{-4}$. 
        \end{itemize}

   \item Next, we define the bridge box to distinguish between pixels that sample the bridge and the clusters regions, just like in Fig. \ref{fig:ShapS_y_RMs}. Given that we matched the exact same relative geometry between the clusters and groups of the system in the density cube, we can easily define where the vertices of the bridge box are in terms of pixels. We will denote the region delimited by the bridge box by $\mathscr{B}$.

   \item Combining these two we have:
    \begin{itemize}
        \item Bridge: pixels that satisfy $n_{e,\text{proj}}\in \mathscr{N}_{e,\text{Bridge}}$ and inside the bridge box ($\Omega_{\text{Bridge}} \equiv \mathscr{N}_{e,\text{Bridge}}\cap\mathscr{B}$)
        
        \item Clusters: pixels that satisfy $n_{e,\text{proj}}\in \mathscr{N}_{e,\text{Clusters}}$ and outside the box ($\Omega_{\text{Clusters}} \equiv \mathscr{N}_{e,\text{Clusters}}\setminus\mathscr{B}$). 
    \end{itemize}
\end{itemize}

Then, to compute the excess $\RM$ scatter in these regions we perform a Monte-Carlo and bootstrapping method:
\begin{itemize}
    \item First, we draw randomly generated pixels from the mock $\RM$ map. The number of these pixels is set to match the observed areal density of the observed $\RM$s in the on and off-target regions ($\mathcal{A}$), i.e., $\sim36\,\RM$ deg$^{-2}$.
    
    \item These pixel locations represent lines of sight through the full system. Thus, to extract $\sigma_{\RM}^{\Omega}$ in the desired $\Omega$ region we select only the random pixels that satisfy the corresponding definitions: $(x,y)\in \Omega$, and evaluate the rotation measure map at them $\RM(x,y)$ and compute its MAD standard deviation ($\sigma_{\RM}$).
    
    \item This process is done $10^6$ times, which yields a distribution for $\sigma_{\RM}^{\Omega}$. We quote the median of this distribution as the estimate of the scatter and the $68\%$ confidence level interval sets the lower/upper error bars from the $16\%$/$84\%$ quantiles, respectively.
\end{itemize}

This bootstrapping was made to fully sample the SSC region we have modeled, restricted to the same areal density of $\RM$s from our observations. The ASKAP-POSSUM $\RM$s used in this work only represent a sample through this Shapley regions, while with our mock maps we can learn more about them through its statistical properties. The results are in Sect. \ref{sect:miro_signals}.

\subsubsection{$\mathfrak{S}_{\RM}(d_{\text{nrst}})$ profiles and Bayesian model selection}\label{sect:methods_miro_profiles}

In order to constrain magnetic field models obtained from MiRò we computed $\mathfrak{S}_{\RM}(d_{\text{nrst}})$ profiles to directly compare with the observed profile from Sect. \ref{sect:obs_RM_scatter_prof}. In contrast with Eqn. \eqref{eq:scatter_prof_formula}, no measurement error or off-target contribution corrections are needed in this case. Thus, computing these profiles reduces to computing the MAD standard deviation of the mock $\RM$ maps as a function of $d_{\text{nrst}}$. We remind the reader that we created a density cube with the exact same relative distances between the clusters and groups at the plane of the sky. As a consequence, we can directly compare the $d_{\text{nrst}}$ values from the mock maps with those computed from the observations. 

For the observational profile we used the data of the on-target sub-sample. This motivates the decision to use the pixels inside the $\mathscr{N}_{e,\text{Clusters}}$ region of the maps, which was defined to explicitly reproduce the region inside $y_{\text{bdry}}$, to compute $\mathfrak{S}_{\RM}(d_{\text{nrst}})$. We used a sliding window with a fixed step size of $\Delta d_{\text{nrst}}=10^{-2}$ that corresponds to a physical size of $\sim 10\,\mathrm{kpc}$ for an average $r_{500}$ of $\simeq 1\,\mathrm{Mpc}$. The 68$\%$ C.L. error region was obtained from a bootstrap resampling with $10^4$ resamples. 
The resulting profiles can be found in Sect. \ref{sect:scatter_profs_miro_obs_comparison}. \\

To statistically quantify which model best matches the data from the $\mathfrak{S}_{\RM}$ profiles we have defined a Gaussian likelihood function ($\mathcal{L}_k$) for each model $k$ (see Eqn. \eqref{eq:likelihood}).
$N=11$ is the number of bins used for the model selection based on these profiles. For this comparison, we evaluated the mock $\RM$ maps at the exact locations of the polarized sources, and derived $\mathfrak{S}_{\RM}(d_{\text{nrst}})$ profiles with the same sliding window used for the observed profile (Sect. \ref{sect:obs_RM_scatter_prof}). Then, we only used bins in distance that satisfy $d_{\text{nrst}}\leq 0.7$ to compare only in the region where the profile is not dominated by the overlapping structures.  
Given the fact that the error bars are asymmetric, we took into account the upper (lower) error bars whenever $\mathfrak{S}_{\RM,i}^{\text{MiRò}}>\mathfrak{S}_{\RM,i}^{\text{Obs}}$, ($\mathfrak{S}_{\RM,i}^{\text{MiRò}}<\mathfrak{S}_{\RM,i}^{\text{Obs}}$).

To assess the performance of the models in this Bayesian framework, we computed the Bayesian evidence for each model $k$, given the data $d$ which we can determine as 
\begin{equation}
\small{
    p(d|k) = \int \mathcal{L}_k(\Vec{\theta}) p(\Vec{\theta},k) \,d\Vec{\theta}= \int \mathcal{L}_k(\Vec{\theta})\, \delta(\Vec{\theta}-\Vec{\theta}_k) \,d\Vec{\theta} = \mathcal{L}_k(\Vec{\theta}_k),
}
\end{equation}
where we have assumed a fixed parameter space $\Vec{\theta}_k$ (as it is the case for us). Therefore, the Bayesian evidence is just the likelihood for a given model (defined by Eqn. \eqref{eq:likelihood}). Then, the model with the largest $\ln{\mathcal{L}_k}$ is the model with the highest probability of being true given the data. To quantify the performance of the model with highest evidence (labeled by $j$) over the rest of the models ($k$), we compute the Bayes factor $\mathcal{B}_{jk}=\mathcal{L}_j/\mathcal{L}_k$ with respect to it. Following the \citet{kassBayesFactors1995} scale we say that if $\ln{\mathcal{B}_{jk}}\leq 1$ there is ``barely worth mentioning'' evidence in favor of model $j$ with respect to model $k$; if $\ln{\mathcal{B}_{jk}}\leq 3$ there is ``positive'' evidence; if $\ln{\mathcal{B}_{jk}}\leq 5$ there is ``strong'' evidence and if $\ln{\mathcal{B}_{jk}}>5$ there is ``very strong'' evidence.

\subsubsection{Average magnetic field strength}\label{sect:method_B_avg}

We estimated the average magnetic field strength over 2D regions of interest $\Omega$ from the 3D cubes generated by MiRò by
\begin{equation}
    \langle \abs{\Vec{B}} \rangle_{\Omega} = \frac{\sqrt{3}}{N_zN_{\Omega}}\sum_{k=1}^{N_z}\sum_{j=1}^{N_{\Omega}}\abs{ B_{i}(x_j,y_j,z_k) },
    \label{eq:B_avg}
\end{equation}
where we have made use of the Gaussianity of the distribution: $B_x^2 + B_y^2 + B_z^2=3B_i^2$, where this relation holds for any $i=x,y,z$. Here, $N_z=900$ is the number of pixels of the cube along the line of sight, and $N_{\Omega}$ is the number of pixels in the $\Omega$ region. 

\subsection{MiRò $\sigma_{\RM}$ estimates in the clusters and bridge regions}

Using the Monte-Carlo and bootstrap method in Appendix \ref{sect:miro_RM_scatter_method}, we estimated the $\RM$ signature in the bridge and clusters regions as predicted by MiRò. Table \ref{tab:signals_miro} shows these values for the models whose $\sigma^{\text{Clusters}}_{\RM}$ is compatible, within the uncertainties at the 1$\sigma$ level, with the observed value (see Fig. \ref{fig:miro_obs_signal_comparison}). 

\begin{table*}[ht!]
\caption{$\RM$ signatures predicted by MiRò in the bridge and clusters samples computed following Appendix \eqref{sect:miro_RM_scatter_method} for the models that reproduce the observed $\sigma^{\text{Clusters}}_{\RM}$ (see Fig. \ref{fig:miro_obs_signal_comparison}). }
\centering
{\small
\renewcommand{\arraystretch}{1.5}
\setlength{\tabcolsep}{4.5pt}
\begin{tabular}{c|cccccccccc}
($B_{\text{rms}}$[$\mu$G], $\eta$) & (1, 0.5) & (1, 1) & (1.5, 0.5) & (1.5, 1) & (2,0) & (2, 0.5) & (2, 1) & (2.5, 0) & (3, 0) & (3.5, 0) \\ \hline \hline  
$\sigma_{\RM}^{\text{Bridge}}$ [rad/m$^2$] & $16.14^{+6.85}_{-5.40}$ & $15.76^{+7.60}_{-5.45}$
& $24.11^{+10.07}_{-7.99}$ & $23.5^{+11.51}_{-8.24}$ & $15.39^{+5.88}_{-4.93}$ & $32.13^{+13.63}_{-18.18}$ & $30.86^{+14.45}_{-10.67}$ & $19.15^{+7.44}_{-10.67}$ & $23.35^{+9.05}_{-6.2}$ & $27.49^{+19.83}_{-15.49}$ \\ 
$\sigma_{\RM}^{\text{Clusters}}$ [rad/m$^2$] & $18.25^{+4.95}_{-3.98}$ & $17.28^{+5.63}_{-4.14}$
& $27.29^{+7.55}_{-5.98}$ & $25.67^{+8.72}_{-6.29}$ & $18.18^{+4.43}_{-3.84}$ & $37.67^{+10.53}_{-8.42}$ & $35.23^{+12.07}_{-8.67}$ & $22.03^{+5.12}_{-4.45}$ & $25.81^{+6.14}_{-5.29}$ & $31.18^{+7.22}_{-6.32}$

\end{tabular}
}
\label{tab:signals_miro}
\end{table*}

\subsection{RM grid density effect on uncertainties. Prospects with the SKA}\label{sect:RMgrid_dens}

The error bars derived with the Monte-Carlo and bootstrap sampling method (Section \ref{sect:miro_RM_scatter_method}) to compute $\sigma_{\RM}$ in the bridge and clusters regions are large. As an example of this situation, Figure \ref{fig:miro_obs_signal_comparison} illustrates how there is a degeneracy in the predicted $\RM$ scatter in the clusters region between models with very different physical meaning, e.g., $B_{\text{rms}}=1.5\,\mu$G, $\eta=0.5$ and $B_{\text{rms}}=3.5\,\mu$G, $\eta=0$.  

Therefore, we have studied how the relative error on $\sigma_{\RM}^{\text{Bridge}}$ and $\sigma_{\RM}^{\text{Clusters}}$ would be affected by increasing the density of $\RM$s on the sky (Table \ref{tab:density_of_RM_grids}), given that this is one of the main sources of uncertainty in magnetic field estimates we can derive from Faraday Rotation studies \citep{johnsonCharacterizingUncertaintyCluster2020}.

We found that with the current density derived from the ASKAP-POSSUM data in this study ($\rho_{\RM}\simeq 36\,\RM$s/deg$^2$), the error bars are of the order of $\sim25$-$37\%$, for the clusters and bridge regions, respectively. Even for a density of $100\,\RM/\mathrm{deg^2}$, about $\sim 3$ times higher, which is close to the expected density for the SKA \citep{healdMagnetismScienceSquare2020}, we would still obtain a relative error of $\sim15$-$21\%$. To get errors lower than $10\%$, we would have to increase the density of sources to an unrealistic $500\,\RM/\mathrm
{deg^2}$. Note that, given the large angular size and the redshift of the SSC, these estimates serve as a lower limit in the uncertainties for the expectations of similar studies of objects of smaller angular size and at higher redshifts. This result indicates that the uncertainties we can expect for single object studies with $\RM$ grids, will remain of the order of $10\%$ or greater in the next decades as far their density is concerned.

\begin{table}[h]
\caption{Impact on the inherent uncertainties that a statistical analysis from an RM grid has when increasing the density of the grid. }
    \centering
    \renewcommand{\arraystretch}{1.35}
    \begin{tabular}{c|cccc}
        $\rho_{\RM}=\RM\text{s}/\mathrm{deg^2}$ & 36 & 100 & 200 & 500 \\ 
        \hline \hline 
        $(\Delta\sigma/\sigma)^{\text{Bridge}}_{\RM}$ & $37\%$ & $21\%$ & $15\%$ & $9\%$  \\ 
        $(\Delta\sigma/\sigma)^{\text{Clusters}}_{\RM}$ & $25\%$ & $15\%$ & $10\%$ & $6\%$ 
    \end{tabular}
    \tablefoot{The first column (with $36/\mathrm{deg^2}$) corresponds to the current ASKAP-POSSUM observations used in this work. The second column (with $100/\mathrm{deg^2}$) would correspond to the expectations from the SKA \citep{healdMagnetismScienceSquare2020}. The relative errors are the mean of all MiRò models used in our analysis.}
    \label{tab:density_of_RM_grids}
\end{table}

These limitations serve as motivation to do further analysis with the $\RM$ scatter profiles with distance and using other statistics to distinguish between models, such as the Bayesian model selection analysis performed in Sections \ref{sect:scatter_profs_miro_obs_comparison} and \ref{sect:slow_scatter_profiles}.

\subsection{$\mathfrak{S}_{\RM}(n_e)$ profiles from MiRò}\label{sect:RM_scatter_profiles_n_e}

We computed $\RM$ scatter profiles as a function of electron density. For the observational profile, the electron density was estimated using Eqn. \eqref{eq:n_e_obs}, with $L=1.8\,\mathrm{Mpc}$ (Appendix \ref{sect:L}), temperature $T_e=5\times 10^7\,\mathrm{K}$ (Appendix \ref{sect:obs_n_e_degeneracy}), and the values of the $y$-map at the location of the on-target sources. The sliding window and the bootstrapping used was the same as in Sect. \ref{sect:obs_RM_scatter_prof}. 
For the profiles derived from the mock $\RM$ MiRò maps, we used their values evaluated at the pixels inside the $\mathscr{N}_{e,\text{Clusters}}$ region. The electron density values were obtained from this same region of the projected density map $n_{e,\text{proj}}$ (Eqn. \eqref{eq:proj_n_e}). We used a sliding window of fixed size $\Delta n_e=10^{-5}$ cm$^{-3}$ and bootstrap with $10^4$ resamples for the errors.

\begin{table}[h]
\caption{ Bayesian evidence, Bayes factors and K\&R scale result for the MiRò model selection analysis. } 
    \centering
    \small{
    \renewcommand{\arraystretch}{1.1}
    \begin{tabular}{c|ccccc}
        ($B_{\text{rms}},\,\eta$) & (2.5, 0) & ($1$, $0.5$) & ($1.5$, $0.5$) & ($3$, $0$) & ($2$, $0$) \\
        \hline \hline
        $\ln{\mathcal{L}_k}$ & -55.3 & -55.6 & -57.0 & -57.5 & -57.6 \\
        $\ln{\mathcal{B}}$ & ... & 0.3 & 1.8 & 2.3 & 2.4 \\
        K\&R & ... & Barely & Positive & Positive & Positive
    \end{tabular}
    \tablefoot{We compared the value of the $\mathfrak{S}^{\text{MiRò}}_{\RM}(n_e)$ profiles at the position of the on-target sources in the range $n_e \geq 2.5\times10^{-4}$cm$^{-3}$. More details can be found in Appendix \ref{sect:methods_miro_profiles}. $B_{\text{rms}}$ is in $\,\mu\mathrm{G}$.}
\label{tab:bayes_miro_n_e}
}
\end{table}

\begin{figure}[ht!]
    \centering
    \includegraphics[width=1\linewidth,clip=true,trim=0cm 0.8cm 0cm 2.68cm]{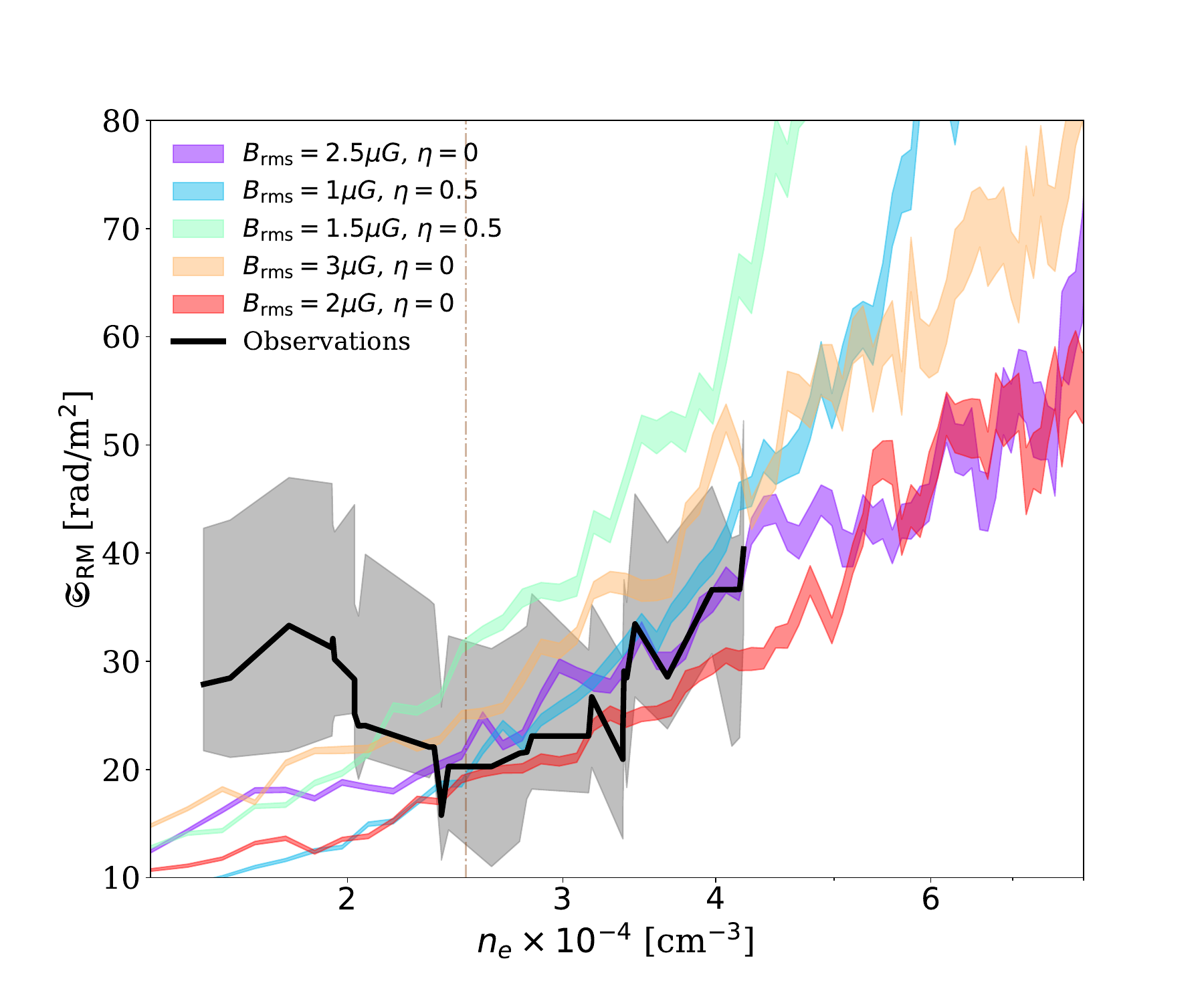}
    \caption{ $\mathfrak{S}_{\RM}(n_e)$ profiles. The color-shaded regions represent the $68\%$ C.L. error regions for the five MiRò profiles with the best performance according to the Bayesian analysis. The black line represents the observed scatter profile and the gray shaded region represents its $68\%$ C.L. error region. The vertical dash-dotted line marks $n_e \geq 2.5\times10^{-4}$cm$^{-3}$, which is the region where overlapping structures do not dominate.}
    
    \label{fig:RM_n_e_scatter_profiles}
\end{figure}

We also carried out a Bayesian model selection analysis to determine the predictive performance of the models, and compare between them following the same procedure as in \ref{sect:methods_miro_profiles}. 
Analogously, we used Eqn. \eqref{eq:likelihood} but the profiles are now binned in electron density values instead. We compared the data with the models only using the values of the mock $\RM$ and $n_{e,\text{proj}}$ maps at the position of the on-target sources and the same sliding window and bootstrap parameters used for the observational profile. We restricted the electron density bins to those that satisfy $n_e \geq 2.5\times10^{-4}$cm$^{-3}$ to avoid overlapping structures dominating the comparison.

Table \ref{tab:bayes_miro_n_e} shows the Bayesian model selection analysis results. The model with highest Bayesian evidence which is an indicator for best agreement between the model and the data has parameters $B_{\text{rms}}=2.5\,\mu\mathrm{G}$, $\eta=0$, which is the same one found from the analysis carried out in Section \ref{sect:scatter_profs_miro_obs_comparison}. The model with $B_{\text{rms}}=1\,\mu\mathrm{G}$, $\eta=0.5$ is the one with the second best performance, and there is only barely worth mentioning evidence against it. 

\subsection{SLOW simulations}\label{sect:slow_methods}

SLOW \citep{Dolag2023} uses initial conditions constructed from the \texttt{CosmicFlows-2} dataset \citep{Tully2013}, to closely resemble the observed local Universe at $z=0$, using the realization number 8 of CLONES \citep[see][and references therin for details on the construction process]{Sorce2018}.
The SLOW simulation produces a close match to observed local galaxy clusters \citep{Hernandez2024} as well as local super clusters, like the Shapley Supercluster \citep[see discussion in][]{Seidel2024}.
For this work we performed a high-resolution zoom-in re-simulation of the Shapley Supercluster region based on initial conditions constructed from the collapse volumes of these structures (Seidel et al., in prep). We used the same physical setup as the \textsc{Coma} zoom simulation presented in \citet{bossSimulatingLOcalWeb2024}, performed with the cosmological \textsc{Tree-SPH} code \textsc{OpenGadget3} code \citep{Groth2023}.
This simulation is a non-radiative MHD simulation, using the MHD solver described in \citet{Dolag2009}, the non-ideal MHD extension by \citet{Bonafede2011}, constrained hyperbolic divergence cleaning \cite[][Steinwandel \& Price; in prep]{Tricco2016} as well as higher order smoothing kernels and improvements to the hydro solver summarized in \citet{Beck2016}.

\subsubsection{Magnetic field amplification mechanisms}\label{sect:slow_amp_mech}

\begin{figure*}[ht!]
\centering
\includegraphics[width=1\linewidth]{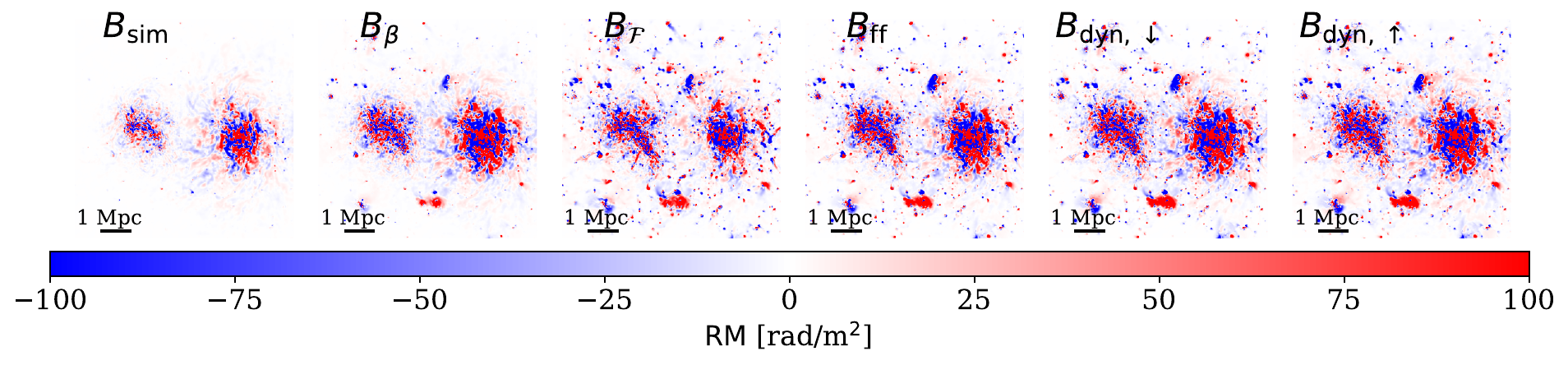}
\caption{Rotation Measure maps of all SLOW magnetic field amplification mechanisms used in this work.}
\label{fig:RM_slow_maps}
\end{figure*}

In this simulation, a seed magnetic field of strength $B=10^{-14}\,\mathrm{G}$ at redshift $z=120$ was set up. This field evolves and is amplified by adiabatic compression and dynamo processes. On the one hand, these mechanisms are enough to provide magnetic field strengths of $\sim \mu\mathrm{G}$ in the cluster centers, on the other hand, the magnetic field strengths in filaments are negligible, at the current resolution. To further amplify the magnetic field in the filaments, several mechanisms are implemented: two of them scaling with pressure, and three with density:
\begin{itemize}
    \item Plasma-$\beta$: magnetic field pressure and thermal pressure are related by a constant factor $B_{\beta}=\sqrt{8\pi P_{th}/\beta}$, with $\beta=20$.
    \item Turbulent velocity: the magnetic field strength is a fraction $\mathcal{F}$ of the turbulent pressure $B_{\mathcal{F}}=\mathcal{F} \sqrt{4\pi \rho v_{\text{turb}}^2}$ \footnote{Note that $\rho\propto n_e$ in this simulation.}. The fraction was set to $\mathcal{F}=1$ assuming pressure equilibrium between turbulent and magnetic pressures\footnote{Turbulent velocities are obtained by performing a Helmholtz-decomposition using the the code \textsc{vortex-p} \citep{vortexp}.}. 
    \item Flux-freezing: from the conservation of the magnetic flux, amplification by 3D compression implies $B_{\text{ff}}\propto n_e^{2/3}$.
    \item Turbulent dynamo $\downarrow$: this model assumes that turbulent dynamo amplifies the field down to densities of $n_e\sim 10^{-4}\,\mathrm{cm^{-3}}$. For lower densities than that, a fitting was made to match \citet{carrettiMagneticFieldEvolution2023} observations of $\sim 30\,\mathrm{nG}$ fields in filaments down to densities of $n_e\sim 10^{-5}\,\mathrm{cm^{-3}}$. The scaling behavior is $B_{\text{dyn,}\downarrow}\propto n_e^{1/2}$.
    \item Turbulent dynamo $\uparrow$: pure saturated dynamo with the same $B_{\text{dyn,}\uparrow}\propto n_e^{1/2}$ scaling.
\end{itemize}
For more details see \citet{bossSimulatingLOcalWeb2024}.
Please note that the models $B_\mathrm{ff}, B_{\text{dyn,}\downarrow}$ and $B_{\text{dyn,}\downarrow}$ are up-scaled by a factor of two in respect to the previous work for a better match with observations.
To model the RM signal we rescale the parallel component of the simulated magnetic field according to the different magnetic field models.

In order to compare the results from the SLOW analog with the observations, we applied very similar techniques to those used to compare with the outputs from MiRò (Appendix \ref{sect:miro_methods}).
The $\RM$ maps for the six different magnetic field amplification mechanisms implemented by SLOW can be see in Fig. \ref{fig:RM_slow_maps}. 

\subsubsection{Excess RM scatter in the Shapley analog}\label{sect:slow_method_RM_scatter}

To estimate the excess $\RM$ scatter in 2D regions ($\sigma_{\RM}^{\Omega}$), we based the analysis on \ref{sect:miro_RM_scatter_method}. This is, we defined the region $\Omega$ in terms of an appropriate electron density range and in terms of a bridge box. 
Regarding density, the boundaries are\begin{itemize}
    \item $\mathscr{N}_{e,\text{Bridge}}\equiv 4\times 10^{-5} \leq n_{e} \, \text{\small [cm$^{-3}$] } \leq 4.5\times 10^{-4}$
    \item $\mathscr{N}_{e,\text{Clusters}}\equiv 4\times 10^{-5} \leq n_{e} \, \text{\small [cm$^{-3}$] } \leq 8\times 10^{-4}$. 
\end{itemize}
This choices were made trying to probe a similar range of densities as in the observations, avoiding lines of sight that go directly through the clusters and groups centers. 
The region $\Omega$ is also defined in terms of a bridge box, whose orientation is parallel to the line that joins the centers of the massive clusters. However, the geometry of the system is not directly comparable to that in the real Shapley, since the positions of the analog SC groups are not exactly the same. In fact, the projected distance between the clusters is a factor $0.8$ lower. Despite this fact, the definition of the bridge box ensures that the groups lie inside it, so that the scatter of the $\RM$ in the bridge region accounts for their contribution. 

In summary, the bridge and clusters regions defined in the SLOW maps are: Bridge: pixels that satisfy $n_{e}\in \mathscr{N}_{e,\text{Bridge}}$ and inside the bridge box; Clusters: pixels that satisfy $n_{e}\in \mathscr{N}_{e,\text{Clusters}}$ and outside the box. 
Once these regions are defined, to obtain $\sigma_{\RM}^{\Omega}$ the same Monte-Carlo and bootstrap methods in Appendix \ref{sect:miro_RM_scatter_method} were used.

\subsubsection{$\mathfrak{S}_{\RM}(d_{\text{nrst}})$ profiles with SLOW}\label{sect:slow_method_scatter_profiles}

We also computed profiles of the $\RM$ scatter with respect to distance. Given that the SLOW SSC analog does not have the groups in the same locations as in the real Shapley, the distance $d_{\text{nrst}}$ was defined as the distance to the nearest cluster only, without considering the distances to the groups. Therefore, we also made a profile from the observations considering the distance to the clusters only. To compare the SLOW profiles with the observational one, we have re-scaled the latter by a factor $\times 0.8$ that corresponds to the ratio between the projected distances between the A3558 and A3562 clusters of the SLOW analog of Shapley, and the real one ($\sim3.5\,\mathrm{Mpc}/4.2\,\mathrm{Mpc}$)

To produce these profiles, we have proceeded in a similar way to Appendix \ref{sect:methods_miro_profiles}. Therefore, we have employed a sliding window approach with a fixed step size of $\Delta d_{\text{nrst}}=0.01$ and bootstrap resampling with $10^4$ resamples. We derived profiles for the $\RM$s derived from all amplification mechanisms in the region $\mathscr{N}_{e,\text{Clusters}}$. Nevertheless, to define the likelihood function for the Bayesian model selection analysis, we have binned these profiles according to the binning of the observational profile. This was done by finding the values of $\mathfrak{S}_{\RM}^{\text{SLOW}}$ at the binned values of $d_{\text{nrst},i}$ from $\mathfrak{S}_{\RM}^{\text{Obs}}$. Then, the likelihood function and the model selection analysis is completely analogous to Appendix \ref{sect:methods_miro_profiles}. The results can be found in Sect. \ref{sect:slow_scatter_profiles}.

\subsection{SLOW $\sigma_{\RM}$ estimates in the clusters and bridge regions}

Table \ref{tab:slow_scatters} shows the excess $\RM$ scatters in the bridge and clusters regions for all SLOW amplification mechanisms for the magnetic field computed using the method in Appendix \ref{sect:slow_method_RM_scatter}.

\begin{table}[ht!]
\caption{$\sigma_{\RM}$ (in rad/m$^2$) for all SLOW magnetic field amplification mechanisms.} 
\centering
{\tiny
\renewcommand{\arraystretch}{1.4}
\setlength{\tabcolsep}{1.5pt}
\begin{tabular}{c|cccccc}
Mechanism & $B_{\text{sim}}$ & $B_{\beta}$ & $B_{\mathcal{F}}$ & $B_{\text{ff}}$ & $B_{\text{dyn}, \downarrow}$ & $B_{\text{dyn}, \uparrow}$ \\ \hline \hline  
$\sigma_{\RM}^{\text{Bridge}}$
& $10.26^{+3.96}_{-3.17}$ & $21.33^{+8.85}_{-6.77}$ & $15.15^{+6.98}_{-4.97}$ & $18.32^{+8.20}_{-5.97}$ & $27.01^{+11.70}_{-8.67}$ & $28.53^{+11.91}_{-9.02}$ \\ 
$\sigma_{\RM}^{\text{Clusters}}$
& $8.04^{+1.89}_{-1.55}$ & $18.85^{+4.51}_{-3.60}$ & $20.48^{+4.50}_{-3.70}$ & $21.08^{+5.10}_{-3.95}$ & $27.01^{+11.70}_{-8.67}$ & $31.19^{+6.78}_{-5.48}$
\end{tabular}}
\label{tab:slow_scatters}
\end{table}

\end{appendix}

\end{document}